\documentclass[aps,prb,twocolumn,showpacs,superscriptaddress,amsmath,amssymb,longbibliography]{revtex4-1}
\usepackage[pdftex]{graphicx}
\usepackage{xcolor}
\usepackage[pdftex,bookmarks=true,bookmarksopen,bookmarksnumbered,
                colorlinks,
                linkcolor=blue,
                citecolor=blue]{hyperref}
\usepackage{bm}
\usepackage{mathtools}
\usepackage{graphicx}
\usepackage{graphics}
\usepackage{dcolumn}
\usepackage{float}
\usepackage{epsfig}
\usepackage{epstopdf} 
\usepackage{color}
\usepackage{amsmath}
\usepackage{textcomp}
\usepackage{multirow}
\usepackage{array}
\usepackage{newfloat}

\DeclareFloatingEnvironment[
    fileext=los,
    name=Figure,
    placement=tbhp,
    within=none,
]{suppfig}

\renewcommand{\figurename}{Figure}
\makeatletter
\renewcommand*{\fnum@figure}{{\normalfont\bfseries \figurename~\thefigure}}
\renewcommand*{\@caption@fignum@sep}{ $|$}
\newcommand*{\rom}[1]{\expandafter\@slowromancap\romannumeral #1@}
\makeatother

\newcommand {\ket} [1] {|{ #1 \rangle}}
\newcommand {\bra} [1] {{\langle #1 }|}

\pdfpageattr {/Group << /S /Transparency /I true /CS /DeviceRGB>>} 

\begin{document}
\title{Fidelity benchmarks for two-qubit gates in silicon}

\author{W. Huang}
\email[wister.huang@student.unsw.edu.au]{}
\affiliation{Center for Quantum Computation and Communication Technology, School of Electrical Engineering and Telecommunications, The University of New South Wales, Sydney, NSW 2052, Australia}

\author{C. H. Yang}
\affiliation{Center for Quantum Computation and Communication Technology, School of Electrical Engineering and Telecommunications, The University of New South Wales, Sydney, NSW 2052, Australia}

\author{K. W. Chan}
\affiliation{Center for Quantum Computation and Communication Technology, School of Electrical Engineering and Telecommunications, The University of New South Wales, Sydney, NSW 2052, Australia}

\author{T. Tanttu}
\affiliation{Center for Quantum Computation and Communication Technology, School of Electrical Engineering and Telecommunications, The University of New South Wales, Sydney, NSW 2052, Australia}

\author{B. Hensen}
\affiliation{Center for Quantum Computation and Communication Technology, School of Electrical Engineering and Telecommunications, The University of New South Wales, Sydney, NSW 2052, Australia}

\author{R. C. C. Leon}
\affiliation{Center for Quantum Computation and Communication Technology, School of Electrical Engineering and Telecommunications, The University of New South Wales, Sydney, NSW 2052, Australia}

\author{M. A. Fogarty}
\altaffiliation[Now at ]{London Centre for Nanotechnology, UCL, 17-19 Gordon St, London WC1H 0AH, United Kingdom}
\affiliation{Center for Quantum Computation and Communication Technology, School of Electrical Engineering and Telecommunications, The University of New South Wales, Sydney, NSW 2052, Australia}

\author{J. C. C. Hwang}
\affiliation{Center for Quantum Computation and Communication Technology, School of Electrical Engineering and Telecommunications, The University of New South Wales, Sydney, NSW 2052, Australia}

\author{F. E. Hudson}
\affiliation{Center for Quantum Computation and Communication Technology, School of Electrical Engineering and Telecommunications, The University of New South Wales, Sydney, NSW 2052, Australia}

\author{K. M. Itoh}
\affiliation{School of Fundamental Science and Technology, Keio University, 3-14-1 Hiyoshi, Kohoku-ku, Yokohama 223-8522, Japan}

\author{A. Morello}
\affiliation{Center for Quantum Computation and Communication Technology, School of Electrical Engineering and Telecommunications, The University of New South Wales, Sydney, NSW 2052, Australia}

\author{A. Laucht}
\affiliation{Center for Quantum Computation and Communication Technology, School of Electrical Engineering and Telecommunications, The University of New South Wales, Sydney, NSW 2052, Australia}

\author{A. S. Dzurak}
\email[a.dzurak@unsw.edu.au]{}
\affiliation{Center for Quantum Computation and Communication Technology, School of Electrical Engineering and Telecommunications, The University of New South Wales, Sydney, NSW 2052, Australia}

\date{\today}

\begin{abstract}
\end{abstract}


\maketitle
\textbf{Universal quantum computation will require qubit technology based on a scalable platform, together with quantum error correction protocols that place strict limits on the maximum infidelities for one- and two-qubit gate operations~\cite{Knill1997,Fowler2012}. While a variety of qubit systems have shown high fidelities at the one-qubit level~\cite{Kok2007,Haffner2008,Barends2014,Rong2015,Muhonen2014,Veldhorst2014,Nichol2017}, superconductor technologies have been the only solid-state qubits manufactured via standard lithographic techniques which have demonstrated two-qubit fidelities near the fault-tolerant threshold~\cite{Barends2014}. Silicon-based quantum dot qubits are also amenable to large-scale manufacture and can achieve high single-qubit gate fidelities (exceeding 99.9~\%) using isotopically enriched silicon \cite{RevModPhys.85.961,itoh_watanabe_2014,Ladd2018}. However, while two-qubit gates have been demonstrated in silicon\cite{Veldhorst2015,Watson2018,Zajac2018}, it has not yet been possible to rigorously assess their fidelities using randomized benchmarking, since this requires sequences of significant numbers of qubit operations ($\gtrsim 20$) to be completed with non-vanishing fidelity. Here, for qubits encoded on the electron spin states of gate-defined quantum dots, we demonstrate Bell state tomography with fidelities ranging from 80~\% to 89~\%, and two-qubit randomized benchmarking with an average Clifford gate fidelity of 94.7~\% and average Controlled-ROT (CROT) fidelity of 98.0~\%. These fidelities are found to be limited by the relatively slow gate times employed here compared with the decoherence times $T_2^*$ of the qubits. Silicon qubit designs employing fast gate operations based on high Rabi frequencies~\cite{Kawakami2014,Takeda2016,Yoneda2018}, together with advanced pulsing techniques~\cite{RevModPhys.76.1037}, should therefore enable significantly higher fidelities in the near future.}

\begin{figure*}
	\includegraphics[width=\linewidth]{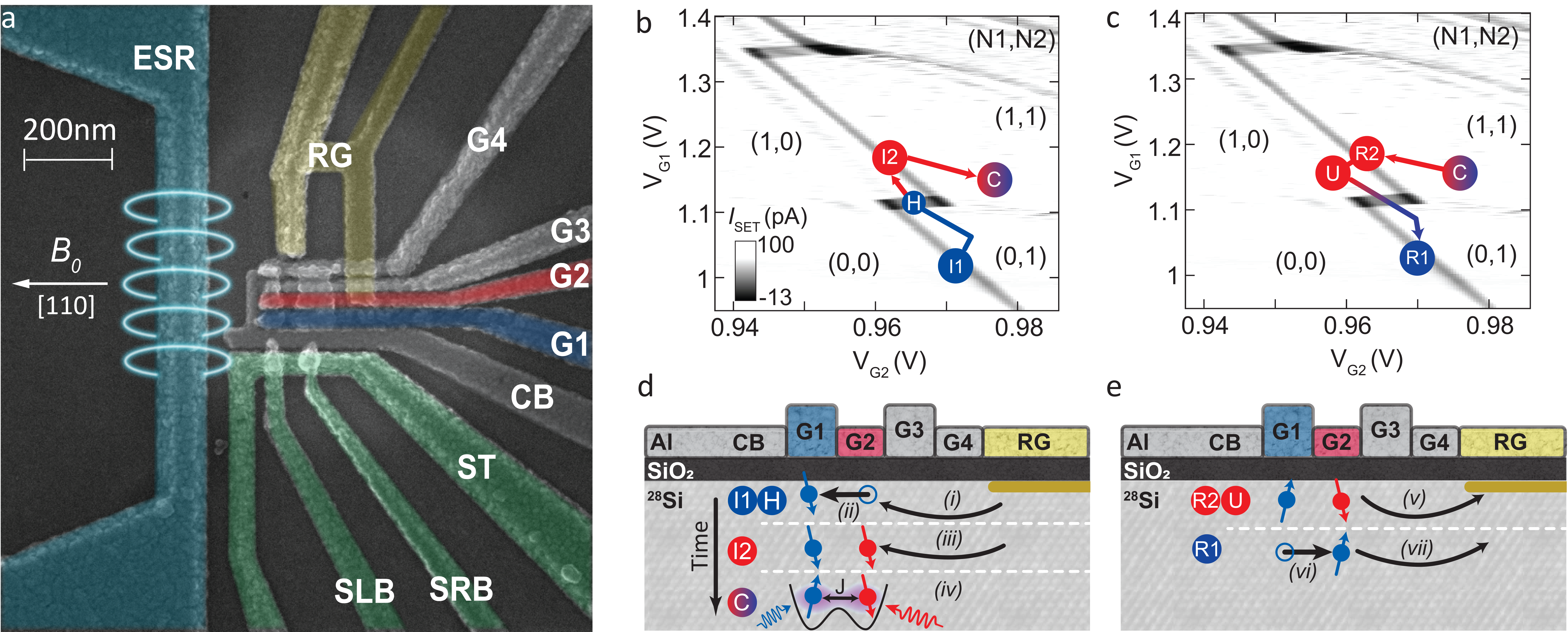}
	\caption{\label{fig:Devicelayout}\textbf{ Two-qubit device layout and operation.}
		\textbf{a}, False colour scanning electron microscope image of the device. Two quantum dots D1 and D2 are formed underneath gates G1 (blue) and G2 (red). The gates CB (purple), G3 and G4 (grey) form confinement barriers that laterally define the quantum dots. RG (yellow) is the reservoir gate that supplies electrons to the quantum dots. The gate electrodes ST, SLB and SRB (green) define a single electron transistor, designed to sense charge movement in the quantum dot region. An AC current running through the ESR line (light blue) generates an oscillating magnetic field to manipulate the electron spins. The direction of the external magnetic field $B_{0}$ is indicated by the white arrow.
		\textbf{b-e}, Control path in the charge stability diagram and schematic depicting initialization and readout: 
		\textit{(i.)} Load a spin-down electron from the reservoir into D2 by biasing to the (0,0)-(0,1) transition (I1) for 2.75~ms. 
		\textit{(ii.)} Move the electron to a spin relaxation hot-spot~\cite{Yang2013} (H) close to the (0,1)-(1,0) anti-crossing and keep it there for $300$~$\mu$s to improve the initialization fidelity~\cite{Watson2018,Zhao2017}. Then transfer the electron to D1 by moving it through the anti-crossing~\cite{Baart2016}, completing the initialization of qubit \textcolor[rgb]{0,0,1}{Q1} as $\ket{\downarrow}$ (I1-H-I2 in the stability diagram).
		\textit{(iii.)} Load another spin-down electron into D2 by biasing to the (1,0)-(1,1) transition (I2) for 2.75~ms to initialize qubit \textcolor[rgb]{1,0,0}{Q2} as $\ket{\downarrow}$. This sequence initializes the system $\ket{\textcolor[rgb]{0,0,1}{Q1},\textcolor[rgb]{1,0,0}{Q2}}$ as $\ket{\downarrow\downarrow}$. The two-qubit system is now ready for operation.
		\textit{(iv.)} Perform single-qubit and two-qubit quantum operations on the two qubits in the (1,1) region (C) by using sequences of selective ESR pulses. 
		\textit{(v.)} Read out the qubit \textcolor[rgb]{1,0,0}{Q2} at the (1,0)-(1,1) transition via spin-dependent tunneling~\cite{Elzerman2004} by biasing to R2 for 2.75~ms, then ensure D2 is unloaded by pulsing deeper into the (1,0) region (U) for 3~ms. 
		\textit{(vi.)} Transfer the qubit \textcolor[rgb]{0,0,1}{Q1} from dot D1 to D2 by adiabatically sweeping through the (1,0)-(0,1) anti-crossing within 5~$\mu$s, which is fast enough to avoid relaxation at the hot spot~\cite{Baart2016}.
		\textit{(vii.)} Read out \textcolor[rgb]{0,0,1}{Q1} at the (0,0)-(0,1) transition (R1) for 2.75~ms. This concludes the operational sequence. In our devices, performing readout by shuttling \textcolor[rgb]{0,0,1}{Q1} from D1 to D2 is advantageous over reading out \textcolor[rgb]{0,0,1}{Q1} at the (1,0)-(0,0) transition directly, due to the slow tunneling rate from D1 to the reservoir.}
\end{figure*}

Silicon provides an ideal environment for spin qubits thanks to its compatibility with industrial manufacturing technologies and the near-perfect nuclear-spin vacuum that isotopically enriched $^{28}$Si provides~\cite{RevModPhys.85.961,itoh_watanabe_2014}. Qubits can be encoded directly on the spins of individual nuclei, donor-bound electrons, or electrons confined in gate-defined quantum dots, or they can be encoded in subspaces provided by two or more spins~\cite{Ladd2018}. Electrostatic gate electrodes allow initialization, readout~\cite{Elzerman2004} and, in some cases, manipulation of qubits~\cite{Petta2180} to be implemented with local electrical pulses. For qubits encoded on single spins, one-qubit gates can be driven using an AC magnetic field to perform electron spin resonance (ESR) directly~\cite{Pla2012,Veldhorst2014}, through an AC electric field produced by a gate electrode combined with the magnetic field gradient from an on-chip micro-magnet~\cite{Pioro2008,Kawakami2014,Takeda2016}, or with an AC electric field acting on the spin-orbit field~\cite{Maurand2016,Huang2017,Corna2018}. In enriched $^{28}$Si devices such one-qubit gates have attained fidelities of 99.9~\% or above~\cite{Muhonen2015,Yoneda2018,Chan2018}.

Two-qubit gates, required to complete the universal gate set, are commonly implemented in spin systems as the $\sqrt{SWAP}$~\cite{Petta2180,Nowack1269}, the C-Phase~\cite{Veldhorst2015,Watson2018} or the CROT~\cite{Veldhorst2015,Zajac2018}. While the $\sqrt{SWAP}$ and the C-Phase gates require fast temporal control of the exchange interaction $J$, accurately synchronized with spin resonance pulses, the CROT can also be implemented with constant $J$~~\cite{Kalra2014}, alleviating the requirements on exchange control and gate electrode bandwidth. Here, in a silicon double quantum dot system, we show how the full two-qubit Clifford gate set can be constructed entirely using ESR pulses in the presence of constant exchange coupling, and use this to perform both Bell state tomography and Clifford-based randomized benchmarking, providing the first detailed analysis of two-qubit gate fidelities in a silicon-based system.

Figure~\ref{fig:Devicelayout}a shows a scanning electron microscope (SEM) image of a silicon-metal-oxide-semiconductor (Si-MOS) double quantum dot device, nominally identical to the one measured and similar to the one that we previously used to demonstrate a two-qubit logic gate~\cite{Veldhorst2015}. The device was fabricated on a natural silicon substrate with a $900$~nm thick isotopically enriched $^{28}$Si epi-layer (residual $^{29}$Si concentration of 800 ppm~\cite{itoh_watanabe_2014}). Aluminium gate electrodes were fabricated using multi-layer gate stack technology~\cite{Angus2007}. Quantum dots D1 and D2 are formed underneath gates G1 and G2, however the exact dot centre positions can be influenced by local strain fields in the device~\cite{Thorbeck2015}. The tunnel rate between the dots and the reservoir RG (yellow) can be modified by adjusting the voltages on G3 and G4 (grey). An external magnetic field $B_{0}=1.42$~T creates a Zeeman splitting of $E_\text{Z}=g \mu_\mathrm{B} B_{0}\approx 0.16$~meV, corresponding to an ESR frequency $f=E_\text{Z}/h=39.33$~GHz, where $g$ is the electron $g$-factor, $\mu_\mathrm{B}$ is the Bohr magneton and $h$ is Planck's constant. When operating the device in a dilution refrigerator at an electron temperature of $T_{e}\approx 100$~mK, the energy gap between spin $\ket{\uparrow}$ and $\ket{\downarrow}$ states allows us to read the electron spin state via spin-dependent tunneling~\cite{Elzerman2004} and selectively load a $\ket{\downarrow}$ electron for initialization. An on-chip ESR antenna (light blue) creates the oscillating magnetic field $B_1$ to perform qubit operations~\cite{Dehollain2012}.

Figures~\ref{fig:Devicelayout}b,c are charge stability diagrams of the double quantum dot system comprising dots D1 and D2, recorded by measuring the current through the single-electron transistor (SET) charge sensor with a double lock-in technique~\cite{Yang2011}. The charge occupancies of D1 and D2 are labeled ($N_1$,$N_2$). Our two-qubit system operates in the sequence schematically depicted in Fig.~\ref{fig:Devicelayout}d,e and described in the figure caption.

\begin{figure*}
	\includegraphics[width=\linewidth]{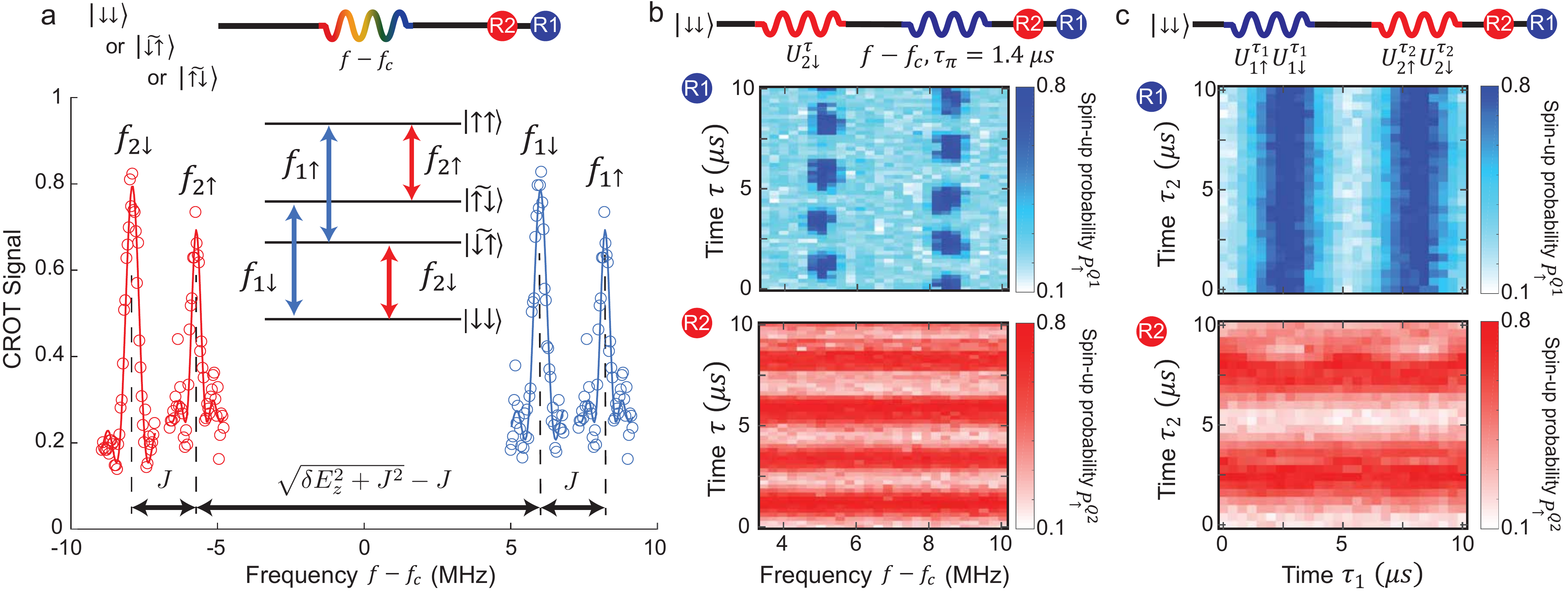}
	\caption{\label{fig:Rabi}\textbf{ Independent and Conditional Two-Qubit Control.}
	\textbf{a} ESR spectra of the two-qubit system. Here, the peaks are power-broadened and the linewidths are given by the respective Rabi frequencies. We prepare $\ket{\textcolor[rgb]{0,0,1}{Q1},\textcolor[rgb]{1,0,0}{Q2}}$ in either $\ket{\downarrow\downarrow}$, $\ket{\tilde{\downarrow\uparrow}}$ or $\ket{\tilde{\uparrow\downarrow}}$ and measure the spin-up probability of \textcolor[rgb]{0,0,1}{Q1} and \textcolor[rgb]{1,0,0}{Q2} as a function of the applied microwave frequency. Four distinct ESR peaks arise due to the presence of a finite exchange coupling $J$ and a Zeeman energy difference $\delta E_\textrm{Z}$, centered around $f_c=\bar{E_Z}/h=39.33$~GHz. Each resonance peak represents a rotation of the target qubit conditional on the state of the control qubit (CROT signal). The energy level diagram (inset) maps each peak to the corresponding transition between a pair of two-qubit eigenstates. 
	\textbf{b} Controlled qubit rotations are naturally implemented by pulsing at individual resonance frequencies. A first pulse $U^{\tau}_{2\downarrow}$ performs Rabi rotations on \textcolor[rgb]{1,0,0}{Q2} that result in the resonance frequency of \textcolor[rgb]{0,0,1}{Q1} oscillating between $f_{1\downarrow}$ and $f_{1\uparrow}$. 
	\textbf{c} Independent qubit control can be achieved under the presence of constant $J$ by applying microwave pulses at the two conditional frequencies simultaneously. A first pulse at $U_{1\uparrow}^{\tau_1}U_{1\downarrow}^{\tau_1}$ defines the state \textcolor[rgb]{0,0,1}{Q1}. A second pulse $U_{2\uparrow}^{\tau_2}U_{2\downarrow}^{\tau_2}$ then rotates \textcolor[rgb]{1,0,0}{Q2}, independent of the state of \textcolor[rgb]{0,0,1}{Q1}.
} 
\end{figure*}

During microwave control, when operating the device deep in the (1,1) charge stability region, the system can be described by a Hamiltonian in a diagonalized basis~\cite{Meunier2011,Kalra2014} $(\uparrow\uparrow,\tilde{\uparrow\downarrow},\tilde{\downarrow\uparrow},\downarrow\downarrow)$:

\begin{equation}
\arraycolsep 0.3 ex
H=\frac{1}{2}\left( \begin{array}{cccc}
2\bar{E_\textrm{Z}} & {\color{red}\gamma_{2\uparrow}}B_1& {\color{blue}\gamma_{1\uparrow}}B_1 & 0 \\
{\color{red}\gamma_{2\uparrow}}B_1^* & \tilde{\delta E_\textrm{Z}}-J & 0 & {\color{blue}\gamma_{1\downarrow}}B_1 \\
{\color{blue}\gamma_{1\uparrow}}B_1^* & 0 &  -\tilde{\delta E_\textrm{Z}}-J & {\color{red}\gamma_{2\downarrow}}B_1 \\
0 &  {\color{blue}\gamma_{1\downarrow}}B_1^* &  {\color{red}\gamma_{2\downarrow}}B_1^* & -2\bar{E_\textrm{Z}} \\
\end{array}\right). 
\end{equation}

Here, $\gamma_{n\downarrow}$ ($\gamma_{n\uparrow}$) is the effective gyromagnetic ratio that couples qubit $n$ to the oscillating magnetic field $B_1$ created by the ESR antenna when the other qubit is in the $\ket{\downarrow}$ ($\ket{\uparrow}$) state, $J$ is the exchange coupling, $\bar{E_\textrm{Z}}$ is the average Zeeman energy, and $\tilde{\delta E_\textrm{Z}}$ is the difference in Zeeman energies. The corresponding energy spectrum is shown in Fig.~\ref{fig:Rabi}a. We extract a difference in Zeeman energy for the two qubits of $\delta E_\textrm{Z}/h=13.26$~MHz at $B_{0}=1.42$~T, which arises from $g$-factor variations due to local electric field gradients and Si/SiO$_2$ interface roughness, mediated by spin-orbit coupling~\cite{Veldhorst2015b}. This splitting is $\sim 500$-times greater than the intrinsic ESR linewidth of $29$~kHz, providing us with independent control over the two qubits. In addition, the exchange coupling $J$ further splits both resonance frequencies, providing us with conditional control of one qubit dependent on the state of the other qubit. $J$ is tunable via gates G1 and G2 (see Figs.~S1,S2), but we keep it constant during our control sequences ($J/h=1.06$~MHz for Bell state tomography and $J/h=1.59$~MHz for randomized benchmarking). We now define $U^{\tau}_{1\uparrow}$ to be a microwave pulse of duration $\tau$ at the frequency that rotates \textcolor[rgb]{0,0,1}{Q1} when \textcolor[rgb]{1,0,0}{Q2} is $\ket{\uparrow}$. The pulses $U^{\tau}_{1\downarrow}$, $U^{\tau}_{2\uparrow}$, $U^{\tau}_{2\downarrow}$ are defined equivalently. 

Since $J$ is non-zero, applying a microwave pulse at a single resonance frequency will lead to a conditional rotation of the target qubit, as we demonstrate in Fig.~\ref{fig:Rabi}b. We apply a first pulse at $U^{\tau}_{2\downarrow}$ to perform Rabi rotations on \textcolor[rgb]{1,0,0}{Q2} (lower panel). When subsequently measuring the ESR spectrum of \textcolor[rgb]{0,0,1}{Q1} by sweeping the microwave frequency $f-f_c$ for a fixed pulse length of $\tau_\pi=1.4$~$\mu$s, we confirm that the resulting resonance frequency of \textcolor[rgb]{0,0,1}{Q1} oscillates between $f_{1\downarrow}$ and $f_{1\uparrow}$, as observed in Fig.~\ref{fig:Rabi}b (upper panel). We calibrate the length of all four resonant pulses to yield $\pi$-rotations to implement Controlled-Rotation (CROT) and Zero-Control-Rotation (Z-CROT) gates when the control qubit is in the 1 ($\ket{\downarrow}$) or 0 ($\ket{\uparrow}$) state, respectively~\cite{Watson2018}. These gates are equivalent to the CNOT and Z-CNOT gates except for an additional phase factor. 

In order to achieve single-qubit control independent of the state of the other qubit, we need to apply a two-frequency resonance pulse (e.g. $U_{1\uparrow}U_{1\downarrow}$), which yields a $X/2$ gate for a $\frac{\pi}{2}$-rotation. Fig.~\ref{fig:Rabi}c shows the implementation of independent control in the experiment. A two-frequency microwave pulse $U_{1\uparrow}^{\tau_1}U_{1\downarrow}^{\tau_1}$ addressing \textcolor[rgb]{0,0,1}{Q1} is followed by another two-frequency pulse $U_{2\uparrow}^{\tau_2}U_{2\downarrow}^{\tau_2}$ addressing \textcolor[rgb]{1,0,0}{Q2} to demonstrate that the Rabi oscillations of \textcolor[rgb]{1,0,0}{Q2} (lower panel) do not depend on the state of \textcolor[rgb]{0,0,1}{Q1} (upper panel). 

We achieve two-axis control over both qubits by implementing $\pi$-rotations around the z-axis as virtual-Z ($Z_V$) gates~\cite{PhysRevA.96.022330}, which are changes in the phase of the reference frame defined by the multi-level rotating frame (See Supplementary Section III). 
We further characterize the qubit properties by measuring the coherence times in the (1,1) regime to be $T_{2,\textcolor[rgb]{0,0,1}{Q1}}^*=24.3\pm2$~$\mu$s, $T_{2,\textcolor[rgb]{1,0,0}{Q2}}^*=10.5\pm1$~$\mu$s, $T_{2,\textcolor[rgb]{0,0,1}{Q1}}^{\textrm{Hahn}}=290\pm40$~$\mu$s, and $T_{2,\textcolor[rgb]{1,0,0}{Q2}}^{\textrm{Hahn}}=33\pm5$~$\mu$s. (See Extended Data Fig.~\ref{fig:T2data}).

\begin{figure*}
	\includegraphics[width=\linewidth]{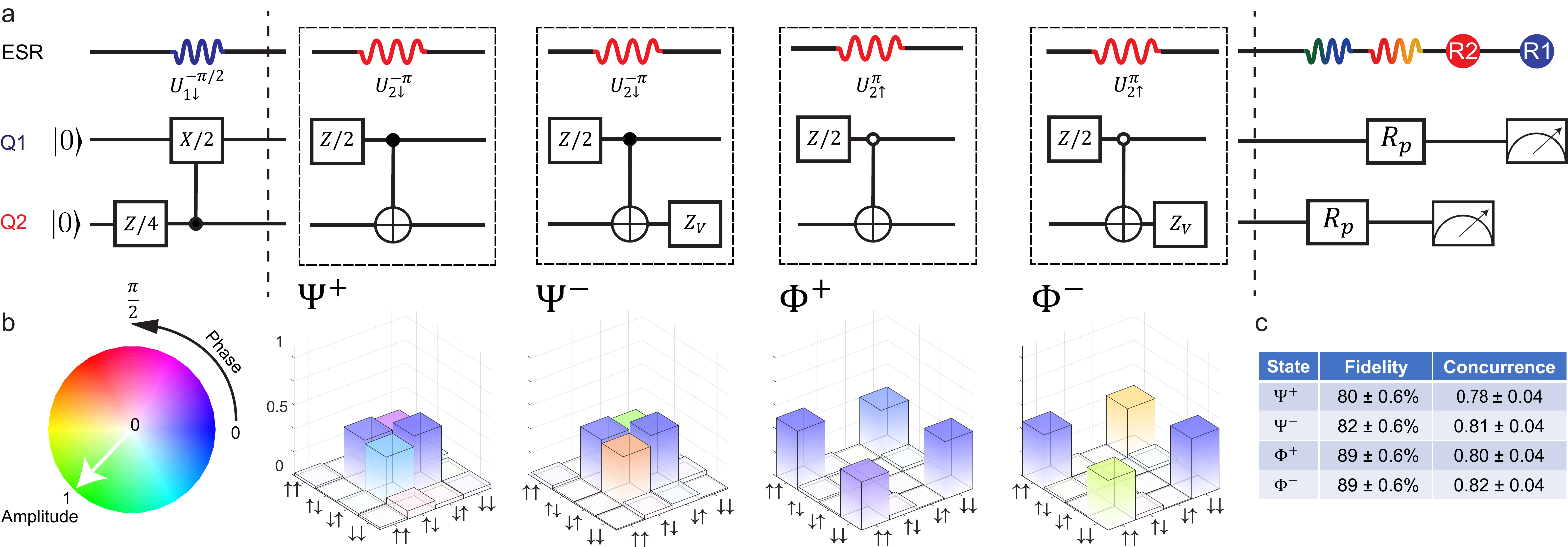}
	\caption{\label{fig:Tomography}\textbf{ Bell State Tomography.} 
	\textbf{a} Pulse sequences to create and measure the four Bell states and their equivalent quantum circuits. The Bell states are created by a single qubit $X_{\pi/2}$ gate followed by a CROT gate ($\ket{\Phi^{-}}$,$\ket{\Phi^{+}}$) or Z-CROT gate ($\ket{\Psi^{-}}$,$\ket{\Psi^{+}}$) that includes a $Z_V$ gate for $\ket{\Psi^{-}}$ and $\ket{\Phi^{-}}$. One of four pre-measurement rotations $Rp=\{I,X/2,-X/2,Y/2,-Y/2\}$ is performed to project the each qubit into the Z, Y, -Y, -X, and X bases to reconstruct the density matrices. 
	\textbf{b} Quantum state tomography of the Bell states. The height of the bars represents the absolute value of the density matrix elements after readout error correction. The phase information is encoded in the colour.
	\textbf{c} Table of Bell state fidelities and concurrences.
	}
\end{figure*}

We continue by performing quantum state tomography on Bell states to demonstrate the creation of entangled states in our two-qubit system and to provide an initial estimate of obtainable gate fidelities (see Methods). We show the quantum circuits and corresponding pulse sequences in Fig.~\ref{fig:Tomography}a. To prepare the four Bell states $\ket{\Phi^{\pm}}=\frac{1}{\sqrt{2}}(\ket{\uparrow\uparrow}\pm\ket{\downarrow\downarrow})$ and $\ket{\Psi^{\pm}}=\frac{1}{\sqrt{2}}(\ket{\tilde{\uparrow\downarrow}}\pm\ket{\tilde{\downarrow\uparrow}})$, we start with the system initialized in the $\ket{\downarrow\downarrow}$ state. We then perform a zero-conditional-$X/2$ pulse on \textcolor[rgb]{0,0,1}{Q1} to bring the system into the $\frac{1}{\sqrt{2}}(\ket{\tilde{\uparrow\downarrow}}+\ket{\downarrow\downarrow})$ state. A CROT or Z-CROT gate is applied to entangle the two qubits. By varying the phase of the underlying ESR pulse to perform either a $\pi$- or $-\pi$-rotation, we include the additional $Z_V$ phase gate on \textcolor[rgb]{0,0,1}{Q1} that is needed to create $\ket{\Phi^{-}}$ and $\ket{\Psi^{-}}$. This results in the creation of the four Bell states. After this sequence we perform one of four pre-measurement rotations $R_p=\{I,X/2,-X/2,Y/2,-Y/2\}$ to achieve projective measurements in the Z, Y, -Y, -X, and X bases, respectively. Although the projection outcome on the -X and -Y basis contains redundant information, it is useful to cancel out systematic errors. The two-qubit density matrix $\rho$ is reconstructed from the combined 25 projection axes with 800 repetitions using maximum likelihood estimation (MLE). We further exclude readout errors by taking readout visibility into account~\cite{Watson2018}. The resulting density matrices are presented in Fig.~\ref{fig:Tomography}b. We calculate the fidelities by comparing the reconstructed states to the ideal Bell state $\ket{\psi}$ using $F={\langle\psi} |\rho |{\psi\rangle}$. The extracted Bell state fidelities demonstrate the creation of highly-entangled states with $F=80-89$~\% and concurrences between 0.78 and 0.82 (see table in Fig.~\ref{fig:Tomography}c).

\begin{figure}
	\includegraphics[width=\linewidth]{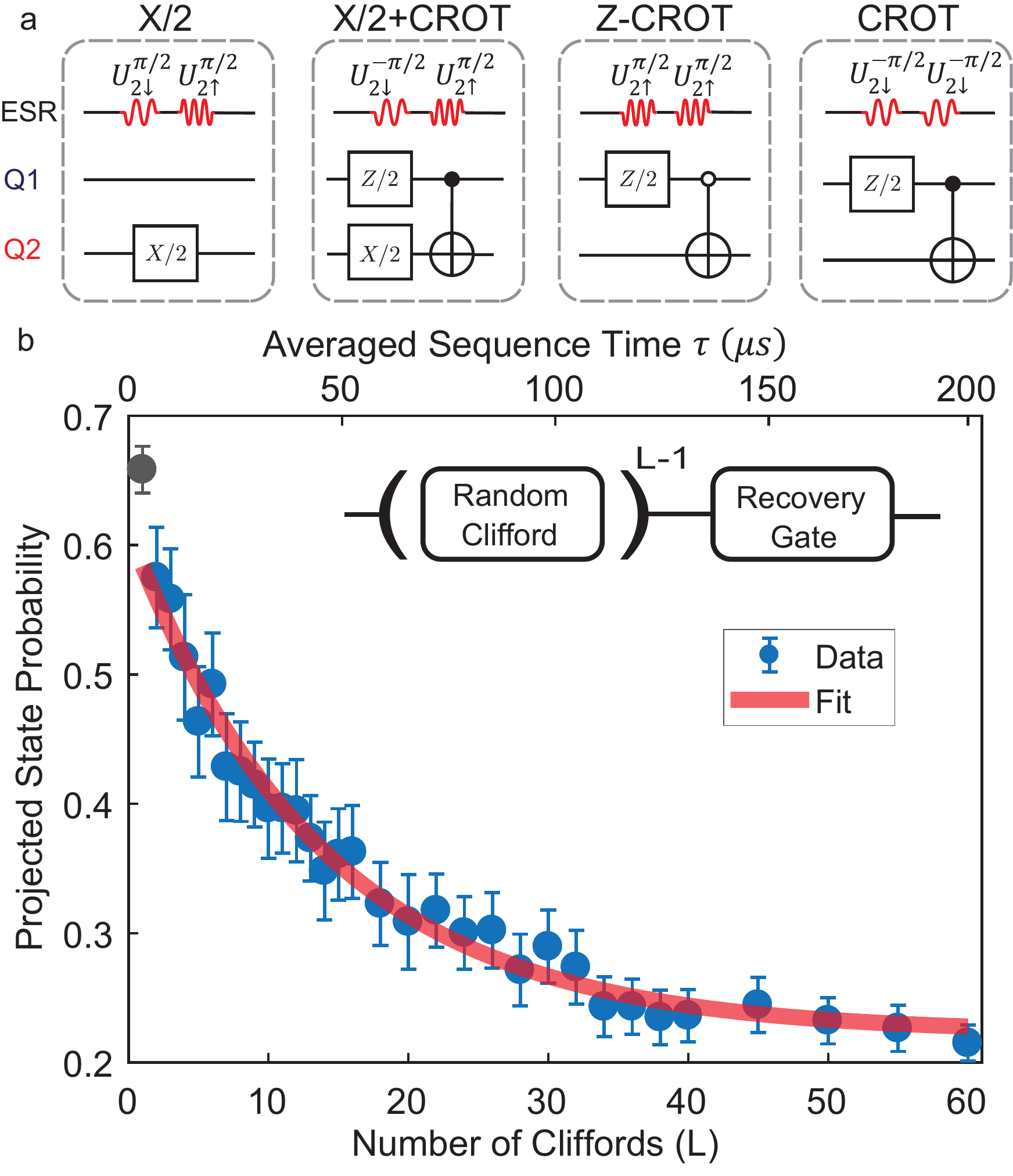}
	\caption{\label{fig:RBM}\textbf{ Two-qubit randomized benchmarking.}
	\textbf{a} Primitive gates $X/2$, $X/2+$CROT, Z-CROT and CROT acting on \textcolor[rgb]{1,0,0}{Q2} and their corresponding ESR pulse sequences. Together with the virtual gate $Z_V/2$ and the gates acting on \textcolor[rgb]{0,0,1}{Q1}, these gates span the two-qubit Clifford space. 
	\textbf{b} Projected state probability as a function of the number of Clifford gates in each sequence. Each sequence is repeated 125 times and the measurement averages over 51 sequences of the same length. The Clifford gates are randomly chosen from 11520 elements of the two-qubit Clifford group, with the $L$-th gate projecting the state to the $\ket{\uparrow\uparrow}$ state. The extracted Clifford fidelity is $F_{\mathrm{Clifford}}=94.7\pm0.8$~\%, the primitive gate fidelity is $F_{\mathrm{primitive}}=98.0\pm0.3$~\%, and the conditional $\frac{\pi}{2}$-pulse fidelity is $F_{\frac{\pi}{2}}^{\mathrm{cond}}=99.0\pm0.15$~\%.
		}
\end{figure}

Tomographic characterization of quantum gates, such as Bell state tomography, is convenient to implement as it requires comparatively short sequences of pulses (see e.g. Fig.~\ref{fig:Tomography}a). It produces a first estimate of the fidelities in the system, however disentangling gate errors from state preparation and measurement (SPAM) errors can be rather imprecise, making the quantification of gate fidelities $>$99~\% almost impossible. Furthermore, it only provides the gate fidelity of a specific gate operation at a time, making it cumbersome to sample over the whole two-qubit Clifford space. Randomized benchmarking (RB) protocols, on the other hand, are inherently insensitive to SPAM errors and allow characterization of the average gate fidelity with much higher accuracy. This is because gates are repeated many times, gate errors accumulate, and RB measures only the decay of the process fidelity as a function of gate operations.

In Fig.~\ref{fig:RBM} we show Clifford-based two-qubit RB of our system. We are using the primitive gates $X/2$, $X/2+$CROT, Z-CROT and CROT and the virtual gate $Z_V/2$ to construct the two-qubit Clifford space~\cite{PhysRevA.96.022330} (see Methods). The primitive gates acting on \textcolor[rgb]{1,0,0}{Q2} are shown in Fig.~\ref{fig:RBM}a. We implement the $X/2$ gate as two sequential $\frac{\pi}{2}$-pulses $U_{2\downarrow}^{\pi/2}U_{2\uparrow}^{\pi/2}$, and the CROT (Z-CROT) gate as two sequential $\frac{\pi}{2}$-pulses at the same frequency $U_{2\downarrow}^{\pi/2}U_{2\downarrow}^{\pi/2}$ ($U_{2\uparrow}^{\pi/2}U_{2\uparrow}^{\pi/2}$), with all pulses being optimized to reduce crosstalk (see Supp. Inf. Sec.~III). All 11520 gates of the two-qubit Clifford space can be generated by sequences of 1-4 primitive gates (see Extended Data Table~\ref{table:Cliffordtable}), and our implementation results in an average of 2.57 primitive gates and 5.14 $\frac{\pi}{2}$-pulses per Clifford gate. As the virtual gate $Z_V/2$ is performed by instantaneous phase switching on the microwave source and does not include any physical pulses that interact with the qubit system, we do not include $Z_V/2$ in the gate counts.

The RB protocol randomly generates a gate sequence of varying length $L-1$ with all gates chosen from the two-qubit Clifford group. A final $L$-th Clifford gate is appended at the end of each sequence to project the final state to $\ket{\uparrow\uparrow}$ and is chosen randomly out of the possible gates that give the required projection. Fig.~\ref{fig:RBM}b shows the result of the projected state probability for $L$ Clifford gates being applied to the initial state $\ket{\downarrow\downarrow}$. We fit the decay with the function $P=A(1-\frac{4}{3}r_c)^L+B$, but do not include the $L=1$ data point as this gate is not a random element of the whole two-qubit Clifford set. The fitting parameters $A$ and $B$ absorb the SPAM errors, leaving $r_c$ as the error per Clifford gate. We obtain a Clifford gate fidelity of $F_{\mathrm{Clifford}}=1-r_c=94.7\pm0.8$~\%, a primitive gate fidelity of $F_{\mathrm{primitive}}=98.0\pm 0.3$~\%, and a conditional $\frac{\pi}{2}$-pulse fidelity of $F_{\frac{\pi}{2}}^{\mathrm{cond}}=99.0\pm 0.15$~\%. As all primitive gates are very similar in construction, we expect the fidelity of the entangling CROT gate to be very close to the average primitive gate fidelity. 

Two-qubit RB is much more sensitive to decoherence than single-qubit RB (see Extended Data Fig.~\ref{fig:T2data}, \ref{table:RBcharge}). In single-qubit RB, the qubit is almost continuously driven around the Bloch sphere, which somewhat refocuses fluctuations in the precession frequency~\cite{Ryan2009}, while the coherent drive also makes the qubit less sensitive to noise~\cite{Laucht2017}. In our mode of operation with constant $J$, the qubits sit idle for approximately 50~\% of the two-qubit RB sequence, making them susceptible to dephasing on a timescale of $T_2^*$. As a consequence, the projected state probability decays on a comparable timescale (see top axis in Fig.~\ref{fig:RBM}b). Faster gate operations would allow more gates to be completed within $T_2^*$~\cite{Kawakami2014,Takeda2016}, however the comparatively small value of $J/h=1.59$~MHz limits our utilizable Rabi frequencies due to power-broadening of the excitation profile. Possible remedies include higher $J$ coupling, optimized shaped pulses that reduce accidental excitation of neighbouring transitions~\cite{RevModPhys.76.1037}, dynamical decoupling of the qubits during the idle times, and samples with higher isotopic purification. Over the 13 hours of data acquisition used to compile the data in Fig.~\ref{fig:RBM}b, we used frequency feedback to compensate for drifts and jumps of the ESR frequencies caused by magnetic field decay, local charge fluctuations and residual $^{29}$Si nuclear spins (see Extended Data Fig.~\ref{fig:Freqtrackprotocal}).
More sophisticated frequency tracking schemes could also contribute to higher fidelities~\cite{Sergeevich2011,Shulman2014,Delbecq2016}.

In conclusion, we have shown that the full two-qubit Clifford gate set can be constructed purely using magnetic resonance pulses acting on silicon spin qubits, and have used this to obtain the two-qubit gate fidelity using randomized benchmarking. This technique, which utilizes a constant exchange coupling between qubits, provides a convenient way of benchmarking fidelities without the need for complex synchronization between exchange gate and spin resonance pulses. The two qubits can be controllably entangled, as demonstrated by the generation of the four Bell states with fidelities of $F=80-89$~\% and concurrences between 0.78 and 0.82. We measured a platform-independent two-qubit gate fidelity of $F_{\mathrm{Clifford}}=94.7\pm0.8$~\%, which translates into a conditional $\frac{\pi}{2}$-pulse fidelity of $F_{\frac{\pi}{2}}^{\mathrm{cond}}=99.0\pm0.15$~\% and $F_{\mathrm{primitive}}=98.0\pm0.3$~\% for the primitive gates that include the CROT. We identify that the main source of infidelity in our experiment is the slow Rabi frequency ($\sim410$~kHz) in comparison with the dephasing rate. While barely affecting $T_2^*$, Rabi frequencies as high as 30~MHz have recently been demonstrated using electric-dipole spin resonance techniques in silicon devices~\cite{Yoneda2018}. Two-qubit fidelities reaching the required limits for fault-tolerance~\cite{Fowler2012} are therefore within reach and underpin silicon as a technology platform with good prospects for scalability to the large numbers of qubits needed for universal quantum computing~\cite{Vandersypen2017,Veldhorst2017}.

\subsection*{Acknowledgments}
We thank S. Bartlett, R. Harper, L. M. K. Vandersypen, T. D. Ladd, and N. C. Jones for insightful discussions.
	We acknowledge support from the	US Army Research Office (W911NF-13-1-0024 and W911NF-17-1-0198), the Australian Research Council (CE11E0001017), and the NSW Node of the Australian National Fabrication Facility. 
	The views and conclusions contained in this document are those of the authors and should not be interpreted as representing the official policies, either expressed or implied, of the Army Research Office or the U.S. Government. 
	The U.S. Government is authorized to reproduce and distribute reprints for Government purposes notwithstanding any copyright notation herein. B.H. acknowledges support from the Netherlands Organization for Scientific Research (NWO) through a Rubicon Grant. K.M.I. acknowledges support from a Grant-in-Aid for Scientific Research by MEXT, NanoQuine, FIRST, and the JSPS Core-to-Core Program.

\clearpage
\renewcommand\thesuppfig{\textbf{\arabic{suppfig}}}
\renewcommand{\suppfigname}{\textbf{Extended Data Figure}}
\renewcommand{\tablename}{\textbf{Extended Data Table}}
\renewcommand\thetable{\textbf{\Roman{table}}}

\section*{Methods}
\subsection*{Experimental setup}
The measurements were conducted in an Oxford Instruments wet dilution refrigerator with base temperature $T_{bath}\approx$ 30~mK and electron temperature $T_{electron}\approx$ 100~mK. DC voltages were applied using battery-powered voltage sources (Stanford Research Systems SIM928) and added to voltage pulses produced with an arbitrary waveform generator (LeCroy ArbStudio 1104 AWG) through resistive voltage dividers/combiners. The voltages applied to the device are attenuated 1:5 for DC voltages and 1:25 for voltage pulses. Low pass filters were included for slow and fast lines (10~Hz to 80~MHz). ESR pulses were delivered by an Agilent E8257D microwave vector signal generator and attenuated at the 1.5 K
stage (10 dB) and the 20 mK stage (3 dB). The internal AWG of the vector signal generator is used to perform IQ modulation. The stability diagrams are obtained using a double lock-in technique (Stanford Research Systems SR830) with dynamic voltage compensation~\cite{Yang2011}.

\subsection*{State tomography}
\label{section:Tomography}
The two qubit density operator can be described by $\rho=\frac{1}{2^n}\sum_{i,j}S_{ij} \sigma_i \otimes\sigma_j$. Where $\sigma$ are Pauli matrices $\sigma\in\{I,X,Y,Z\}$. Pre-measurement rotations $R_p=\{I,X/2,-X/2,Y/2,-Y/2\}$ are performed on both qubits to project the state into the $\{Z,Y,-Y,-X,X\}\otimes\{Z,Y,-Y,-X,X\}$ basis. The readout error for each probability set $P=(P_{\uparrow\uparrow},P_{\uparrow\downarrow},P_{\downarrow\uparrow},P_{\downarrow\downarrow})$ is corrected~\cite{Watson2018} by $P=(F_1 \otimes F_2)^{-1} P_{Measured}$, where
\begin{equation}
	F_i=\begin{pmatrix}
	F_{i \uparrow} & 1-F_{i \downarrow} \\
	1-F_{i \uparrow} & F_{i \downarrow}
	\end{pmatrix}.
\end{equation}

In the experiment, we characterized the readout fidelity using the amplitude of the Rabi oscillations and obtained $F_{1\uparrow}=0.83,F_{1\downarrow}=0.92,F_{2\uparrow}=0.84,F_{2\downarrow}=0.94$.

Maximum likelihood estimation (MLE) is used to estimate the inversion of the matrix. The density matrix is firstly restricted to be non-negative Hermitian:
\begin{equation}
\hat\rho=\frac{T^{\dagger}T}{Tr(T^{\dagger}T)},
\end{equation}

where the division by $Tr(T^{\dagger}T)$ is to ensure normalization. Assuming the measurement error of each qubit state follows a Gaussian distribution, it is possible to estimate the closest density matrix to the measured state. The matrix $T$ for the two qubit system can be parameterized by 15 independent parameters $t_1,....,t_{15}$. The resulting matrix $\rho$ is the closest estimation of the real density matrix by minimizing the following cost function

\begin{equation}
L(t_1,t_2,...,t_n)=\sum_v \frac{(\bra{\psi_v}\rho(t_1,t_2,...,t_n)\ket{{\psi_v}}-n_v)^2}{2\bra{\psi_v} \rho(t_1,t_2,...,t_n)\ket{\psi_v}}.
\end{equation}

\begin{table}[b]
	\begin{center}
		\begin{tabular}{ p{2cm}||p{5cm}}
			\hline
			$L_{primitive}$ & Number of Clifford gates \\
			\hline \hline
			0 & 16 \\
			\hline
			1 & 384 \\
			\hline
			2 & 4176 \\
			\hline
			3 & 6912 \\
			\hline
			4 & 32 \\
			\hline
		\end{tabular}
	\caption{ Number of Clifford gates built from each primitive gate length.}
	\label{table:Cliffordtable}
	\end{center}
\end{table}

\subsection*{Generating Clifford gates}
The Clifford group consists of all elements $C$ that fulfill the condition  $C^\dagger P C\in\pm P$, where P are the Pauli matrices. The Clifford gates in our experiment are generated by different combinations of the primitive gates described in Fig.~\ref{fig:RBM}a and a virtual-$Z_V/2$ gate on each qubit. All primitive gates consist of two conditional $\frac{\pi}{2}$-pulses on the same qubit, and we adjusted the pulse amplitude to ensure all conditional $\frac{\pi}{2}$-pulses have the same length of $0.61$~$\mu$s. We then generate the Clifford group by computer search. Comparing all possible combinations of primitive gates to the gates in the Clifford group, we find the combinations that require the minimal numbers of primitive gates $L_{primitive}$. The number of Clifford gates that can be produced by sequencing $L_{primitive}$ gates is summarized in Table~\ref{table:Cliffordtable}. All two-qubit Clifford gates can be built out of 4 primitive gates, and on average each Clifford gate is composed of 2.5694 primitive gates. In two-qubit RB experiments, the projected state probability is fitted to $P=A(1-4/3r_c)+B$, where A and B are free parameters that absorb SPAM errors. The average Clifford gate fidelity is calculated as $F_c=1-r_c$ and the primitive gate fidelity is $F_{primitive}=1-r_c/2.5694$.

\begin{suppfig*}
	\includegraphics[width=\linewidth]{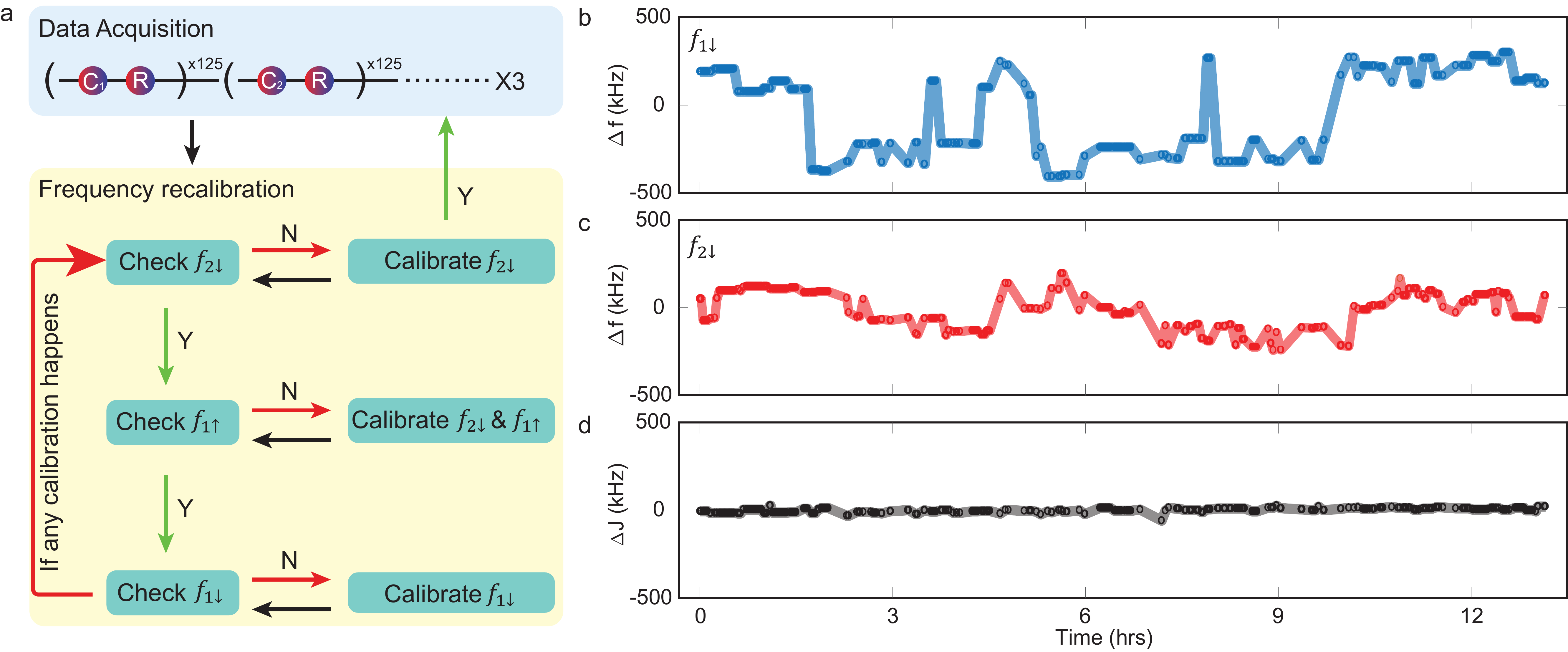}
	\caption{\label{fig:Freqtrackprotocal}\textbf{ Frequency tracking protocol}
	\textbf{(a)} Frequency calibration of the ESR frequencies is implemented by interleaving calibration sequences with the RB experiment. After acquisition of 3 random sequences (1 sequence is repeated 125 times), we check if the ESR frequency is still on resonance by applying a low-power (26~dB lower than the typical operating power) $\pi$-rotation. If the spin-up probability is above the threshold of 50~\% of the readout visibility, the experiment will continue. If the spin-up probability is below the threshold, the resonance frequency will be recalibrated until all ESR frequencies pass the check, and the measurement will continue. 
	\textbf{(b,c)} Resonance frequency fluctuations $\Delta f=f_{1\downarrow}-f_{avg}$ of $f_{1\downarrow}$ (b) and $\Delta f=f_{2\downarrow}-f_{avg}$ of $f_{2\downarrow}$ (c) during the measurement period. We subtracted the average values of the respective frequencies $f_{avg}$ for better visibility. Over 13 hours of data acquisition, \textcolor[rgb]{0,0,1}{Q1} experiences multiple jumps of $\sim600$~kHz, while the fluctuations of \textcolor[rgb]{1,0,0}{Q2} remain within $\sim300$~kHz. Since the resonance frequency fluctuations of \textcolor[rgb]{0,0,1}{Q1} and \textcolor[rgb]{1,0,0}{Q2} are uncorrelated, we exclude fluctuations of $B_0$ or the microwave reference clock as the cause of the frequency changes.
	\textbf{(d)} Variation of exchange coupling $\Delta J = J - J_{avg}$ during the measurement period. We subtracted the average value $J_{avg}$ for better visibility. The exchange coupling is relatively stable during the experiment. If the frequency fluctuations in (b,c) were to originate from charge noise, it is unlikely that $J$ would remain unaffected. Furthermore, since the Stark shift of \textcolor[rgb]{0,0,1}{Q1} and \textcolor[rgb]{1,0,0}{Q2} is $\approx\pm30$~MHz/V, a 600~kHz jump would require a $\sim20$~mV change of the bias voltage applied to the D1 and D2 gates. Such a change in the electrostatic environment would deteriorate qubit readout via the SET charge sensor, but we noticed no significant change of the readout level during the experiment. On this basis, we further exclude charge noise to be the cause of the frequency changes. We conclude that the frequency jumps are most likely caused by spin flips of residual $^{29}$Si nuclei that locally couple to the quantum dots.} 
\end{suppfig*}

\begin{suppfig*}
	\includegraphics[width=\linewidth]{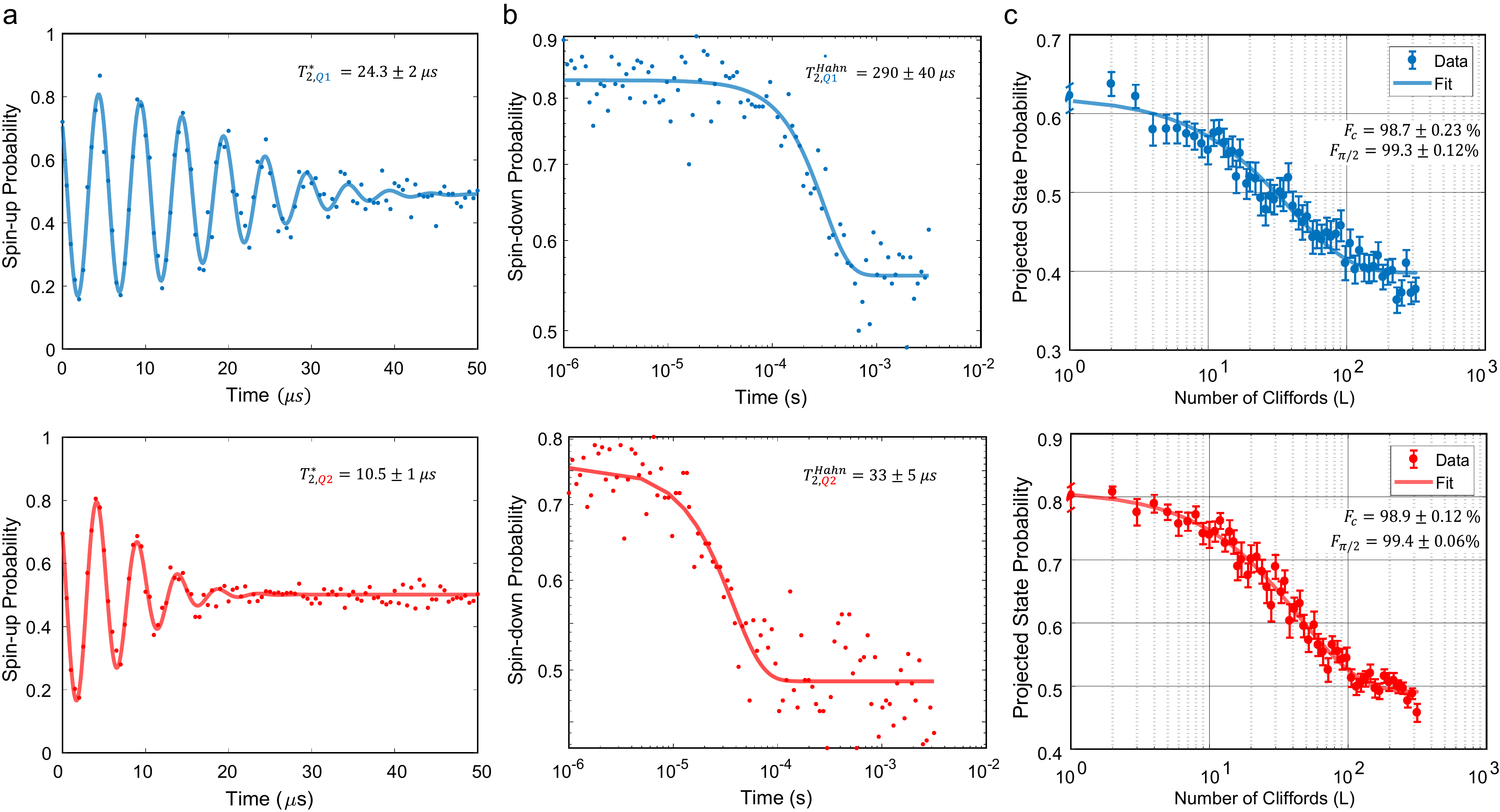}
	\caption{\label{fig:T2data}\textbf{ Single-qubit coherence properties in the (1,1) regime.} Blue data corresponds to {\color{blue} Q1} and red data corresponds to {\color{red} Q2}. We have characterized the single-qubit coherence properties $T_2^*$ and $T_2^{\rm Hahn}$, and measured their control fidelities via single-qubit randomized benchmarking. All data is acquired with the frequency feedback protocol described in Extended Data Fig.~\ref{fig:Freqtrackprotocal}.
		\textbf{a} Spin-up probability as a function of wait time in the Ramsey sequence. $T_{2,{\color{red} Q2}}^*=10.5\pm1$~$\mu$s is much shorter than $T_{2,{\color{blue} Q1}}^*=24.3\pm2$~$\mu$s.
		\textbf{b} Spin-down probability as a function of wait time in the Hahn echo sequence. $T_{2,{\color{red} Q2}}^{\rm Hahn}=33\pm5$~$\mu$s is much shorter than $T_{2,{\color{blue} Q1}}^{\rm Hahn}=290\pm40$~$\mu$s.
		\textbf{c} Single-qubit randomized benchmarking with the other qubit initialized in the $\ket{\downarrow}$ state. Only the frequencies $f_{1\downarrow}$ and $f_{2\downarrow}$ are used for gate operations on {\color{blue} Q1} and {\color{red} Q2} (single tone RB), respectively. The plot shows the projected state probability with increasing number of Clifford gates. The curve is fitted with $P_{\uparrow}=A(1-2r_c)+B$, and the Clifford gate fidelity is given by $F_c=1-r_c$. The single-qubit Clifford gates are on average composed of 1.875 primitive $\pi/2$-pulses, thus the $\pi/2$-pulse fidelity is extracted as $F_{\pi/2}=1-r_c/1.875$. The fidelity for all ESR pulses is in excess of 99~\%.} 
\end{suppfig*}

\begin{table}
	\begin{center}
		\begin{tabular}{ |c|c||c|c|c| } 
			\hline
			Qubit & Charge Regime &$F_{\rm Clifford}$ & $F_{\pi/2}$ \\
			\hline
			\multirow{2}{2em}{{\color{blue} Q1}} & (1,0) & 99.0 $\pm$ 0.38 \% & 99.5 $\pm$ 0.20 \% \\
			\cline{2-4}
			&   (1,1) &  98.7 $\pm$ 0.23 \% & 99.3 $\pm$ 0.12 \%  \\
			\hline
			\multirow{2}{2em}{{\color{red} Q2}} & (0,1) & 99.1 $\pm$ 0.11 \%  & 99.5 $\pm$ 0.06 \% \\
			\cline{2-4}
			&   (1,1) & 98.9 $\pm$ 0.12 \% & 99.4 $\pm$ 0.06 \%  \\
			\hline
		\end{tabular}
	\end{center}
	\caption {\textbf{ Single-qubit properties in different charge regimes.} $F_{\rm Clifford}$ and $F_{\pi/2}$ are similar in the single electron and the (1,1) charge regime, indicating that the dominant source of error is not noise in the exchange coupling $J$.} \label{table:RBcharge}
\end{table}

\clearpage
\bibliography{library}

\begin{thebibliography}{48}%
\makeatletter
\providecommand \@ifxundefined [1]{%
 \@ifx{#1\undefined}
}%
\providecommand \@ifnum [1]{%
 \ifnum #1\expandafter \@firstoftwo
 \else \expandafter \@secondoftwo
 \fi
}%
\providecommand \@ifx [1]{%
 \ifx #1\expandafter \@firstoftwo
 \else \expandafter \@secondoftwo
 \fi
}%
\providecommand \natexlab [1]{#1}%
\providecommand \enquote  [1]{``#1''}%
\providecommand \bibnamefont  [1]{#1}%
\providecommand \bibfnamefont [1]{#1}%
\providecommand \citenamefont [1]{#1}%
\providecommand \href@noop [0]{\@secondoftwo}%
\providecommand \href [0]{\begingroup \@sanitize@url \@href}%
\providecommand \@href[1]{\@@startlink{#1}\@@href}%
\providecommand \@@href[1]{\endgroup#1\@@endlink}%
\providecommand \@sanitize@url [0]{\catcode `\\12\catcode `\$12\catcode
  `\&12\catcode `\#12\catcode `\^12\catcode `\_12\catcode `\%12\relax}%
\providecommand \@@startlink[1]{}%
\providecommand \@@endlink[0]{}%
\providecommand \url  [0]{\begingroup\@sanitize@url \@url }%
\providecommand \@url [1]{\endgroup\@href {#1}{\urlprefix }}%
\providecommand \urlprefix  [0]{URL }%
\providecommand \Eprint [0]{\href }%
\providecommand \doibase [0]{http://dx.doi.org/}%
\providecommand \selectlanguage [0]{\@gobble}%
\providecommand \bibinfo  [0]{\@secondoftwo}%
\providecommand \bibfield  [0]{\@secondoftwo}%
\providecommand \translation [1]{[#1]}%
\providecommand \BibitemOpen [0]{}%
\providecommand \bibitemStop [0]{}%
\providecommand \bibitemNoStop [0]{.\EOS\space}%
\providecommand \EOS [0]{\spacefactor3000\relax}%
\providecommand \BibitemShut  [1]{\csname bibitem#1\endcsname}%
\let\auto@bib@innerbib\@empty
\bibitem [{\citenamefont {Knill}\ and\ \citenamefont
  {Laflamme}(1997)}]{Knill1997}%
  \BibitemOpen
  \bibfield  {author} {\bibinfo {author} {\bibfnamefont {E.}~\bibnamefont
  {Knill}}\ and\ \bibinfo {author} {\bibfnamefont {R.}~\bibnamefont
  {Laflamme}},\ }\href {\doibase 10.1103/PhysRevA.55.900} {\bibfield  {journal}
  {\bibinfo  {journal} {Phys. Rev. A}\ }\textbf {\bibinfo {volume} {55}},\
  \bibinfo {pages} {900} (\bibinfo {year} {1997})}\BibitemShut {NoStop}%
\bibitem [{\citenamefont {Fowler}\ \emph {et~al.}(2012)\citenamefont {Fowler},
  \citenamefont {Mariantoni}, \citenamefont {Martinis},\ and\ \citenamefont
  {Cleland}}]{Fowler2012}%
  \BibitemOpen
  \bibfield  {author} {\bibinfo {author} {\bibfnamefont {A.~G.}\ \bibnamefont
  {Fowler}}, \bibinfo {author} {\bibfnamefont {M.}~\bibnamefont {Mariantoni}},
  \bibinfo {author} {\bibfnamefont {J.~M.}\ \bibnamefont {Martinis}}, \ and\
  \bibinfo {author} {\bibfnamefont {A.~N.}\ \bibnamefont {Cleland}},\ }\href
  {\doibase 10.1103/PhysRevA.86.032324} {\bibfield  {journal} {\bibinfo
  {journal} {Phys. Rev. A}\ }\textbf {\bibinfo {volume} {86}},\ \bibinfo
  {pages} {032324} (\bibinfo {year} {2012})}\BibitemShut {NoStop}%
\bibitem [{\citenamefont {Kok}\ \emph {et~al.}(2007)\citenamefont {Kok},
  \citenamefont {Munro}, \citenamefont {Nemoto}, \citenamefont {Ralph},
  \citenamefont {Dowling},\ and\ \citenamefont {Milburn}}]{Kok2007}%
  \BibitemOpen
  \bibfield  {author} {\bibinfo {author} {\bibfnamefont {P.}~\bibnamefont
  {Kok}}, \bibinfo {author} {\bibfnamefont {W.~J.}\ \bibnamefont {Munro}},
  \bibinfo {author} {\bibfnamefont {K.}~\bibnamefont {Nemoto}}, \bibinfo
  {author} {\bibfnamefont {T.~C.}\ \bibnamefont {Ralph}}, \bibinfo {author}
  {\bibfnamefont {J.~P.}\ \bibnamefont {Dowling}}, \ and\ \bibinfo {author}
  {\bibfnamefont {G.~J.}\ \bibnamefont {Milburn}},\ }\href
  {https://journals.aps.org/rmp/abstract/10.1103/RevModPhys.79.135} {\bibfield
  {journal} {\bibinfo  {journal} {Reviews of Modern Physics}\ }\textbf
  {\bibinfo {volume} {79}},\ \bibinfo {pages} {135} (\bibinfo {year}
  {2007})}\BibitemShut {NoStop}%
\bibitem [{\citenamefont {H{\"a}ffner}\ \emph {et~al.}(2008)\citenamefont
  {H{\"a}ffner}, \citenamefont {Roos},\ and\ \citenamefont
  {Blatt}}]{Haffner2008}%
  \BibitemOpen
  \bibfield  {author} {\bibinfo {author} {\bibfnamefont {H.}~\bibnamefont
  {H{\"a}ffner}}, \bibinfo {author} {\bibfnamefont {C.~F.}\ \bibnamefont
  {Roos}}, \ and\ \bibinfo {author} {\bibfnamefont {R.}~\bibnamefont {Blatt}},\
  }\href {https://www.sciencedirect.com/science/article/pii/S0370157308003463}
  {\bibfield  {journal} {\bibinfo  {journal} {Physics Reports}\ }\textbf
  {\bibinfo {volume} {469}},\ \bibinfo {pages} {155} (\bibinfo {year}
  {2008})}\BibitemShut {NoStop}%
\bibitem [{\citenamefont {Barends}\ \emph {et~al.}(2014)\citenamefont
  {Barends}, \citenamefont {Kelly}, \citenamefont {Megrant}, \citenamefont
  {Veitia}, \citenamefont {Sank}, \citenamefont {Jeffrey}, \citenamefont
  {White}, \citenamefont {Mutus}, \citenamefont {Fowler}, \citenamefont
  {Campbell} \emph {et~al.}}]{Barends2014}%
  \BibitemOpen
  \bibfield  {author} {\bibinfo {author} {\bibfnamefont {R.}~\bibnamefont
  {Barends}}, \bibinfo {author} {\bibfnamefont {J.}~\bibnamefont {Kelly}},
  \bibinfo {author} {\bibfnamefont {A.}~\bibnamefont {Megrant}}, \bibinfo
  {author} {\bibfnamefont {A.}~\bibnamefont {Veitia}}, \bibinfo {author}
  {\bibfnamefont {D.}~\bibnamefont {Sank}}, \bibinfo {author} {\bibfnamefont
  {E.}~\bibnamefont {Jeffrey}}, \bibinfo {author} {\bibfnamefont {T.~C.}\
  \bibnamefont {White}}, \bibinfo {author} {\bibfnamefont {J.}~\bibnamefont
  {Mutus}}, \bibinfo {author} {\bibfnamefont {A.~G.}\ \bibnamefont {Fowler}},
  \bibinfo {author} {\bibfnamefont {B.}~\bibnamefont {Campbell}},  \emph
  {et~al.},\ }\href {https://www.nature.com/articles/nature13171} {\bibfield
  {journal} {\bibinfo  {journal} {Nature}\ }\textbf {\bibinfo {volume} {508}},\
  \bibinfo {pages} {500} (\bibinfo {year} {2014})}\BibitemShut {NoStop}%
\bibitem [{\citenamefont {Rong}\ \emph {et~al.}(2015)\citenamefont {Rong},
  \citenamefont {Geng}, \citenamefont {Shi}, \citenamefont {Liu}, \citenamefont
  {Xu}, \citenamefont {Ma}, \citenamefont {Kong}, \citenamefont {Jiang},
  \citenamefont {Wu},\ and\ \citenamefont {Du}}]{Rong2015}%
  \BibitemOpen
  \bibfield  {author} {\bibinfo {author} {\bibfnamefont {X.}~\bibnamefont
  {Rong}}, \bibinfo {author} {\bibfnamefont {J.}~\bibnamefont {Geng}}, \bibinfo
  {author} {\bibfnamefont {F.}~\bibnamefont {Shi}}, \bibinfo {author}
  {\bibfnamefont {Y.}~\bibnamefont {Liu}}, \bibinfo {author} {\bibfnamefont
  {K.}~\bibnamefont {Xu}}, \bibinfo {author} {\bibfnamefont {W.}~\bibnamefont
  {Ma}}, \bibinfo {author} {\bibfnamefont {F.}~\bibnamefont {Kong}}, \bibinfo
  {author} {\bibfnamefont {Z.}~\bibnamefont {Jiang}}, \bibinfo {author}
  {\bibfnamefont {Y.}~\bibnamefont {Wu}}, \ and\ \bibinfo {author}
  {\bibfnamefont {J.}~\bibnamefont {Du}},\ }\href
  {https://www.nature.com/articles/ncomms9748} {\bibfield  {journal} {\bibinfo
  {journal} {Nature Communications}\ }\textbf {\bibinfo {volume} {6}},\
  \bibinfo {pages} {8748} (\bibinfo {year} {2015})}\BibitemShut {NoStop}%
\bibitem [{\citenamefont {Muhonen}\ \emph {et~al.}(2014)\citenamefont
  {Muhonen}, \citenamefont {Dehollain}, \citenamefont {Laucht}, \citenamefont
  {Hudson}, \citenamefont {Kalra}, \citenamefont {Sekiguchi}, \citenamefont
  {Itoh}, \citenamefont {Jamieson}, \citenamefont {McCallum}, \citenamefont
  {Dzurak},\ and\ \citenamefont {Morello}}]{Muhonen2014}%
  \BibitemOpen
  \bibfield  {author} {\bibinfo {author} {\bibfnamefont {J.~T.}\ \bibnamefont
  {Muhonen}}, \bibinfo {author} {\bibfnamefont {J.~P.}\ \bibnamefont
  {Dehollain}}, \bibinfo {author} {\bibfnamefont {A.}~\bibnamefont {Laucht}},
  \bibinfo {author} {\bibfnamefont {F.~E.}\ \bibnamefont {Hudson}}, \bibinfo
  {author} {\bibfnamefont {R.}~\bibnamefont {Kalra}}, \bibinfo {author}
  {\bibfnamefont {T.}~\bibnamefont {Sekiguchi}}, \bibinfo {author}
  {\bibfnamefont {K.~M.}\ \bibnamefont {Itoh}}, \bibinfo {author}
  {\bibfnamefont {D.~N.}\ \bibnamefont {Jamieson}}, \bibinfo {author}
  {\bibfnamefont {J.~C.}\ \bibnamefont {McCallum}}, \bibinfo {author}
  {\bibfnamefont {A.~S.}\ \bibnamefont {Dzurak}}, \ and\ \bibinfo {author}
  {\bibfnamefont {A.}~\bibnamefont {Morello}},\ }\href
  {http://dx.doi.org/10.1038/nnano.2014.211} {\bibfield  {journal} {\bibinfo
  {journal} {Nature Nanotechnology}\ }\textbf {\bibinfo {volume} {9}},\
  \bibinfo {pages} {986} (\bibinfo {year} {2014})}\BibitemShut {NoStop}%
\bibitem [{\citenamefont {Veldhorst}\ \emph {et~al.}(2014)\citenamefont
  {Veldhorst}, \citenamefont {Hwang}, \citenamefont {Yang}, \citenamefont
  {Leenstra}, \citenamefont {de~Ronde}, \citenamefont {Dehollain},
  \citenamefont {Muhonen}, \citenamefont {Hudson}, \citenamefont {Itoh},
  \citenamefont {Morello},\ and\ \citenamefont {Dzurak}}]{Veldhorst2014}%
  \BibitemOpen
  \bibfield  {author} {\bibinfo {author} {\bibfnamefont {M.}~\bibnamefont
  {Veldhorst}}, \bibinfo {author} {\bibfnamefont {J.~C.~C.}\ \bibnamefont
  {Hwang}}, \bibinfo {author} {\bibfnamefont {C.~H.}\ \bibnamefont {Yang}},
  \bibinfo {author} {\bibfnamefont {A.~W.}\ \bibnamefont {Leenstra}}, \bibinfo
  {author} {\bibfnamefont {B.}~\bibnamefont {de~Ronde}}, \bibinfo {author}
  {\bibfnamefont {J.~P.}\ \bibnamefont {Dehollain}}, \bibinfo {author}
  {\bibfnamefont {J.~T.}\ \bibnamefont {Muhonen}}, \bibinfo {author}
  {\bibfnamefont {F.~E.}\ \bibnamefont {Hudson}}, \bibinfo {author}
  {\bibfnamefont {K.~M.}\ \bibnamefont {Itoh}}, \bibinfo {author}
  {\bibfnamefont {A.}~\bibnamefont {Morello}}, \ and\ \bibinfo {author}
  {\bibfnamefont {A.~S.}\ \bibnamefont {Dzurak}},\ }\href
  {http://dx.doi.org/10.1038/nnano.2014.216} {\bibfield  {journal} {\bibinfo
  {journal} {Nature Nanotechnology}\ }\textbf {\bibinfo {volume} {9}},\
  \bibinfo {pages} {981} (\bibinfo {year} {2014})}\BibitemShut {NoStop}%
\bibitem [{\citenamefont {Nichol}\ \emph {et~al.}(2017)\citenamefont {Nichol},
  \citenamefont {Orona}, \citenamefont {Harvey}, \citenamefont {Fallahi},
  \citenamefont {Gardner}, \citenamefont {Manfra},\ and\ \citenamefont
  {Yacoby}}]{Nichol2017}%
  \BibitemOpen
  \bibfield  {author} {\bibinfo {author} {\bibfnamefont {J.~M.}\ \bibnamefont
  {Nichol}}, \bibinfo {author} {\bibfnamefont {L.~A.}\ \bibnamefont {Orona}},
  \bibinfo {author} {\bibfnamefont {S.~P.}\ \bibnamefont {Harvey}}, \bibinfo
  {author} {\bibfnamefont {S.}~\bibnamefont {Fallahi}}, \bibinfo {author}
  {\bibfnamefont {G.~C.}\ \bibnamefont {Gardner}}, \bibinfo {author}
  {\bibfnamefont {M.~J.}\ \bibnamefont {Manfra}}, \ and\ \bibinfo {author}
  {\bibfnamefont {A.}~\bibnamefont {Yacoby}},\ }\href
  {https://www.nature.com/articles/s41534-016-0003-1} {\bibfield  {journal}
  {\bibinfo  {journal} {npj Quantum Information}\ }\textbf {\bibinfo {volume}
  {3}},\ \bibinfo {pages} {3} (\bibinfo {year} {2017})}\BibitemShut {NoStop}%
\bibitem [{\citenamefont {Zwanenburg}\ \emph {et~al.}(2013)\citenamefont
  {Zwanenburg}, \citenamefont {Dzurak}, \citenamefont {Morello}, \citenamefont
  {Simmons}, \citenamefont {Hollenberg}, \citenamefont {Klimeck}, \citenamefont
  {Rogge}, \citenamefont {Coppersmith},\ and\ \citenamefont
  {Eriksson}}]{RevModPhys.85.961}%
  \BibitemOpen
  \bibfield  {author} {\bibinfo {author} {\bibfnamefont {F.~A.}\ \bibnamefont
  {Zwanenburg}}, \bibinfo {author} {\bibfnamefont {A.~S.}\ \bibnamefont
  {Dzurak}}, \bibinfo {author} {\bibfnamefont {A.}~\bibnamefont {Morello}},
  \bibinfo {author} {\bibfnamefont {M.~Y.}\ \bibnamefont {Simmons}}, \bibinfo
  {author} {\bibfnamefont {L.~C.~L.}\ \bibnamefont {Hollenberg}}, \bibinfo
  {author} {\bibfnamefont {G.}~\bibnamefont {Klimeck}}, \bibinfo {author}
  {\bibfnamefont {S.}~\bibnamefont {Rogge}}, \bibinfo {author} {\bibfnamefont
  {S.~N.}\ \bibnamefont {Coppersmith}}, \ and\ \bibinfo {author} {\bibfnamefont
  {M.~A.}\ \bibnamefont {Eriksson}},\ }\href {\doibase
  10.1103/RevModPhys.85.961} {\bibfield  {journal} {\bibinfo  {journal} {Rev.
  Mod. Phys.}\ }\textbf {\bibinfo {volume} {85}},\ \bibinfo {pages} {961}
  (\bibinfo {year} {2013})}\BibitemShut {NoStop}%
\bibitem [{\citenamefont {Itoh}\ and\ \citenamefont
  {Watanabe}(2014)}]{itoh_watanabe_2014}%
  \BibitemOpen
  \bibfield  {author} {\bibinfo {author} {\bibfnamefont {K.~M.}\ \bibnamefont
  {Itoh}}\ and\ \bibinfo {author} {\bibfnamefont {H.}~\bibnamefont
  {Watanabe}},\ }\href {\doibase 10.1557/mrc.2014.32} {\bibfield  {journal}
  {\bibinfo  {journal} {MRS Communications}\ }\textbf {\bibinfo {volume} {4}},\
  \bibinfo {pages} {143–157} (\bibinfo {year} {2014})}\BibitemShut {NoStop}%
\bibitem [{\citenamefont {Ladd}\ and\ \citenamefont
  {Carroll}(2018)}]{Ladd2018}%
  \BibitemOpen
  \bibfield  {author} {\bibinfo {author} {\bibfnamefont {T.~D.}\ \bibnamefont
  {Ladd}}\ and\ \bibinfo {author} {\bibfnamefont {M.~S.}\ \bibnamefont
  {Carroll}},\ }\href {\doibase
  https://doi.org/10.1016/B978-0-12-803581-8.09736-8} {\bibfield  {journal}
  {\bibinfo  {journal} {Encyclopedia of Modern Optics (Second Edition)}\ ,\
  \bibinfo {pages} {467}} (\bibinfo {year} {2018})}\BibitemShut {NoStop}%
\bibitem [{\citenamefont {Veldhorst}\ \emph
  {et~al.}(2015{\natexlab{a}})\citenamefont {Veldhorst}, \citenamefont {Yang},
  \citenamefont {Hwang}, \citenamefont {Huang}, \citenamefont {Dehollain},
  \citenamefont {Muhonen}, \citenamefont {Simmons}, \citenamefont {Laucht},
  \citenamefont {Hudson}, \citenamefont {Itoh}, \citenamefont {Morello},\ and\
  \citenamefont {Dzurak}}]{Veldhorst2015}%
  \BibitemOpen
  \bibfield  {author} {\bibinfo {author} {\bibfnamefont {M.}~\bibnamefont
  {Veldhorst}}, \bibinfo {author} {\bibfnamefont {C.~H.}\ \bibnamefont {Yang}},
  \bibinfo {author} {\bibfnamefont {J.~C.~C.}\ \bibnamefont {Hwang}}, \bibinfo
  {author} {\bibfnamefont {W.}~\bibnamefont {Huang}}, \bibinfo {author}
  {\bibfnamefont {J.~P.}\ \bibnamefont {Dehollain}}, \bibinfo {author}
  {\bibfnamefont {J.~T.}\ \bibnamefont {Muhonen}}, \bibinfo {author}
  {\bibfnamefont {S.}~\bibnamefont {Simmons}}, \bibinfo {author} {\bibfnamefont
  {A.}~\bibnamefont {Laucht}}, \bibinfo {author} {\bibfnamefont {F.~E.}\
  \bibnamefont {Hudson}}, \bibinfo {author} {\bibfnamefont {K.~M.}\
  \bibnamefont {Itoh}}, \bibinfo {author} {\bibfnamefont {A.}~\bibnamefont
  {Morello}}, \ and\ \bibinfo {author} {\bibfnamefont {A.~S.}\ \bibnamefont
  {Dzurak}},\ }\href {http://dx.doi.org/10.1038/nature15263} {\bibfield
  {journal} {\bibinfo  {journal} {Nature}\ }\textbf {\bibinfo {volume} {526}},\
  \bibinfo {pages} {410} (\bibinfo {year} {2015}{\natexlab{a}})}\BibitemShut
  {NoStop}%
\bibitem [{\citenamefont {Watson}\ \emph {et~al.}(2018)\citenamefont {Watson},
  \citenamefont {Philips}, \citenamefont {Kawakami}, \citenamefont {Ward},
  \citenamefont {Scarlino}, \citenamefont {Veldhorst}, \citenamefont {Savage},
  \citenamefont {Lagally}, \citenamefont {Friesen}, \citenamefont
  {Coppersmith}, \citenamefont {Eriksson},\ and\ \citenamefont
  {Vandersypen}}]{Watson2018}%
  \BibitemOpen
  \bibfield  {author} {\bibinfo {author} {\bibfnamefont {T.~F.}\ \bibnamefont
  {Watson}}, \bibinfo {author} {\bibfnamefont {S.~G.~J.}\ \bibnamefont
  {Philips}}, \bibinfo {author} {\bibfnamefont {E.}~\bibnamefont {Kawakami}},
  \bibinfo {author} {\bibfnamefont {D.~R.}\ \bibnamefont {Ward}}, \bibinfo
  {author} {\bibfnamefont {P.}~\bibnamefont {Scarlino}}, \bibinfo {author}
  {\bibfnamefont {M.}~\bibnamefont {Veldhorst}}, \bibinfo {author}
  {\bibfnamefont {D.~E.}\ \bibnamefont {Savage}}, \bibinfo {author}
  {\bibfnamefont {M.~G.}\ \bibnamefont {Lagally}}, \bibinfo {author}
  {\bibfnamefont {M.}~\bibnamefont {Friesen}}, \bibinfo {author} {\bibfnamefont
  {S.~N.}\ \bibnamefont {Coppersmith}}, \bibinfo {author} {\bibfnamefont
  {M.~A.}\ \bibnamefont {Eriksson}}, \ and\ \bibinfo {author} {\bibfnamefont
  {L.~M.~K.}\ \bibnamefont {Vandersypen}},\ }\href
  {http://dx.doi.org/10.1038/nature25766} {\bibfield  {journal} {\bibinfo
  {journal} {Nature}\ }\textbf {\bibinfo {volume} {555}},\ \bibinfo {pages}
  {633} (\bibinfo {year} {2018})}\BibitemShut {NoStop}%
\bibitem [{\citenamefont {Zajac}\ \emph {et~al.}(2017)\citenamefont {Zajac},
  \citenamefont {Sigillito}, \citenamefont {Russ}, \citenamefont {Borjans},
  \citenamefont {Taylor}, \citenamefont {Burkard},\ and\ \citenamefont
  {Petta}}]{Zajac2018}%
  \BibitemOpen
  \bibfield  {author} {\bibinfo {author} {\bibfnamefont {D.~M.}\ \bibnamefont
  {Zajac}}, \bibinfo {author} {\bibfnamefont {A.~J.}\ \bibnamefont
  {Sigillito}}, \bibinfo {author} {\bibfnamefont {M.}~\bibnamefont {Russ}},
  \bibinfo {author} {\bibfnamefont {F.}~\bibnamefont {Borjans}}, \bibinfo
  {author} {\bibfnamefont {J.~M.}\ \bibnamefont {Taylor}}, \bibinfo {author}
  {\bibfnamefont {G.}~\bibnamefont {Burkard}}, \ and\ \bibinfo {author}
  {\bibfnamefont {J.~R.}\ \bibnamefont {Petta}},\ }\href {\doibase
  10.1126/science.aao5965} {\bibfield  {journal} {\bibinfo  {journal}
  {Science}\ ,\ \bibinfo {pages} {eaao5965}} (\bibinfo {year}
  {2017})}\BibitemShut {NoStop}%
\bibitem [{\citenamefont {Kawakami}\ \emph {et~al.}(2014)\citenamefont
  {Kawakami}, \citenamefont {Scarlino}, \citenamefont {Ward}, \citenamefont
  {Braakman}, \citenamefont {Savage}, \citenamefont {Lagally}, \citenamefont
  {Friesen}, \citenamefont {Coppersmith}, \citenamefont {Eriksson},\ and\
  \citenamefont {Vandersypen}}]{Kawakami2014}%
  \BibitemOpen
  \bibfield  {author} {\bibinfo {author} {\bibfnamefont {E.}~\bibnamefont
  {Kawakami}}, \bibinfo {author} {\bibfnamefont {P.}~\bibnamefont {Scarlino}},
  \bibinfo {author} {\bibfnamefont {D.~R.}\ \bibnamefont {Ward}}, \bibinfo
  {author} {\bibfnamefont {F.~R.}\ \bibnamefont {Braakman}}, \bibinfo {author}
  {\bibfnamefont {D.~E.}\ \bibnamefont {Savage}}, \bibinfo {author}
  {\bibfnamefont {M.~G.}\ \bibnamefont {Lagally}}, \bibinfo {author}
  {\bibfnamefont {M.}~\bibnamefont {Friesen}}, \bibinfo {author} {\bibfnamefont
  {S.~N.}\ \bibnamefont {Coppersmith}}, \bibinfo {author} {\bibfnamefont
  {M.~A.}\ \bibnamefont {Eriksson}}, \ and\ \bibinfo {author} {\bibfnamefont
  {L.~M.~K.}\ \bibnamefont {Vandersypen}},\ }\href
  {http://dx.doi.org/10.1038/nnano.2014.153} {\bibfield  {journal} {\bibinfo
  {journal} {Nature Nanotechnology}\ }\textbf {\bibinfo {volume} {9}},\
  \bibinfo {pages} {666} (\bibinfo {year} {2014})}\BibitemShut {NoStop}%
\bibitem [{\citenamefont {Takeda}\ \emph {et~al.}(2016)\citenamefont {Takeda},
  \citenamefont {Kamioka}, \citenamefont {Otsuka}, \citenamefont {Yoneda},
  \citenamefont {Nakajima}, \citenamefont {Delbecq}, \citenamefont {Amaha},
  \citenamefont {Allison}, \citenamefont {Kodera}, \citenamefont {Oda},\ and\
  \citenamefont {Tarucha}}]{Takeda2016}%
  \BibitemOpen
  \bibfield  {author} {\bibinfo {author} {\bibfnamefont {K.}~\bibnamefont
  {Takeda}}, \bibinfo {author} {\bibfnamefont {J.}~\bibnamefont {Kamioka}},
  \bibinfo {author} {\bibfnamefont {T.}~\bibnamefont {Otsuka}}, \bibinfo
  {author} {\bibfnamefont {J.}~\bibnamefont {Yoneda}}, \bibinfo {author}
  {\bibfnamefont {T.}~\bibnamefont {Nakajima}}, \bibinfo {author}
  {\bibfnamefont {M.~R.}\ \bibnamefont {Delbecq}}, \bibinfo {author}
  {\bibfnamefont {S.}~\bibnamefont {Amaha}}, \bibinfo {author} {\bibfnamefont
  {G.}~\bibnamefont {Allison}}, \bibinfo {author} {\bibfnamefont
  {T.}~\bibnamefont {Kodera}}, \bibinfo {author} {\bibfnamefont
  {S.}~\bibnamefont {Oda}}, \ and\ \bibinfo {author} {\bibfnamefont
  {S.}~\bibnamefont {Tarucha}},\ }\href {\doibase 10.1126/sciadv.1600694}
  {\bibfield  {journal} {\bibinfo  {journal} {Science Advances}\ }\textbf
  {\bibinfo {volume} {2}},\ \bibinfo {pages} {e1600694} (\bibinfo {year}
  {2016})}\BibitemShut {NoStop}%
\bibitem [{\citenamefont {Yoneda}\ \emph {et~al.}(2018)\citenamefont {Yoneda},
  \citenamefont {Takeda}, \citenamefont {Otsuka}, \citenamefont {Nakajima},
  \citenamefont {Delbecq}, \citenamefont {Allison}, \citenamefont {Honda},
  \citenamefont {Kodera}, \citenamefont {Oda}, \citenamefont {Hoshi} \emph
  {et~al.}}]{Yoneda2018}%
  \BibitemOpen
  \bibfield  {author} {\bibinfo {author} {\bibfnamefont {J.}~\bibnamefont
  {Yoneda}}, \bibinfo {author} {\bibfnamefont {K.}~\bibnamefont {Takeda}},
  \bibinfo {author} {\bibfnamefont {T.}~\bibnamefont {Otsuka}}, \bibinfo
  {author} {\bibfnamefont {T.}~\bibnamefont {Nakajima}}, \bibinfo {author}
  {\bibfnamefont {M.~R.}\ \bibnamefont {Delbecq}}, \bibinfo {author}
  {\bibfnamefont {G.}~\bibnamefont {Allison}}, \bibinfo {author} {\bibfnamefont
  {T.}~\bibnamefont {Honda}}, \bibinfo {author} {\bibfnamefont
  {T.}~\bibnamefont {Kodera}}, \bibinfo {author} {\bibfnamefont
  {S.}~\bibnamefont {Oda}}, \bibinfo {author} {\bibfnamefont {Y.}~\bibnamefont
  {Hoshi}},  \emph {et~al.},\ }\href
  {https://www.nature.com/articles/s41565-017-0014-x} {\bibfield  {journal}
  {\bibinfo  {journal} {Nature Nanotechnology}\ }\textbf {\bibinfo {volume}
  {13}},\ \bibinfo {pages} {102} (\bibinfo {year} {2018})}\BibitemShut
  {NoStop}%
\bibitem [{\citenamefont {Vandersypen}\ and\ \citenamefont
  {Chuang}(2005)}]{RevModPhys.76.1037}%
  \BibitemOpen
  \bibfield  {author} {\bibinfo {author} {\bibfnamefont {L.~M.~K.}\
  \bibnamefont {Vandersypen}}\ and\ \bibinfo {author} {\bibfnamefont {I.~L.}\
  \bibnamefont {Chuang}},\ }\href {\doibase 10.1103/RevModPhys.76.1037}
  {\bibfield  {journal} {\bibinfo  {journal} {Rev. Mod. Phys.}\ }\textbf
  {\bibinfo {volume} {76}},\ \bibinfo {pages} {1037} (\bibinfo {year}
  {2005})}\BibitemShut {NoStop}%
\bibitem [{\citenamefont {Yang}\ \emph {et~al.}(2013)\citenamefont {Yang},
  \citenamefont {Rossi}, \citenamefont {Ruskov}, \citenamefont {Lai},
  \citenamefont {Mohiyaddin}, \citenamefont {Lee}, \citenamefont {Tahan},
  \citenamefont {Klimeck}, \citenamefont {Morello},\ and\ \citenamefont
  {Dzurak}}]{Yang2013}%
  \BibitemOpen
  \bibfield  {author} {\bibinfo {author} {\bibfnamefont {C.}~\bibnamefont
  {Yang}}, \bibinfo {author} {\bibfnamefont {A.}~\bibnamefont {Rossi}},
  \bibinfo {author} {\bibfnamefont {R.}~\bibnamefont {Ruskov}}, \bibinfo
  {author} {\bibfnamefont {N.}~\bibnamefont {Lai}}, \bibinfo {author}
  {\bibfnamefont {F.}~\bibnamefont {Mohiyaddin}}, \bibinfo {author}
  {\bibfnamefont {S.}~\bibnamefont {Lee}}, \bibinfo {author} {\bibfnamefont
  {C.}~\bibnamefont {Tahan}}, \bibinfo {author} {\bibfnamefont
  {G.}~\bibnamefont {Klimeck}}, \bibinfo {author} {\bibfnamefont
  {A.}~\bibnamefont {Morello}}, \ and\ \bibinfo {author} {\bibfnamefont
  {A.}~\bibnamefont {Dzurak}},\ }\href
  {https://www.nature.com/articles/ncomms3069} {\bibfield  {journal} {\bibinfo
  {journal} {Nature Communications}\ }\textbf {\bibinfo {volume} {4}},\
  \bibinfo {pages} {2069} (\bibinfo {year} {2013})}\BibitemShut {NoStop}%
\bibitem [{\citenamefont {{Zhao}}\ and\ \citenamefont {{Hu}}(2017)}]{Zhao2017}%
  \BibitemOpen
  \bibfield  {author} {\bibinfo {author} {\bibfnamefont {X.}~\bibnamefont
  {{Zhao}}}\ and\ \bibinfo {author} {\bibfnamefont {X.}~\bibnamefont {{Hu}}},\
  }\href@noop {} {\bibfield  {journal} {\bibinfo  {journal} {ArXiv e-prints}\ }
  (\bibinfo {year} {2017})},\ \Eprint {http://arxiv.org/abs/1707.05217}
  {arXiv:1707.05217} \BibitemShut {NoStop}%
\bibitem [{\citenamefont {Baart}\ \emph {et~al.}(2016)\citenamefont {Baart},
  \citenamefont {Shafiei}, \citenamefont {Fujita}, \citenamefont {Reichl},
  \citenamefont {Wegscheider},\ and\ \citenamefont {Vandersypen}}]{Baart2016}%
  \BibitemOpen
  \bibfield  {author} {\bibinfo {author} {\bibfnamefont {T.~A.}\ \bibnamefont
  {Baart}}, \bibinfo {author} {\bibfnamefont {M.}~\bibnamefont {Shafiei}},
  \bibinfo {author} {\bibfnamefont {T.}~\bibnamefont {Fujita}}, \bibinfo
  {author} {\bibfnamefont {C.}~\bibnamefont {Reichl}}, \bibinfo {author}
  {\bibfnamefont {W.}~\bibnamefont {Wegscheider}}, \ and\ \bibinfo {author}
  {\bibfnamefont {L.~M.~K.}\ \bibnamefont {Vandersypen}},\ }\href
  {http://dx.doi.org/10.1038/nnano.2015.291} {\bibfield  {journal} {\bibinfo
  {journal} {Nature Nanotechnology}\ }\textbf {\bibinfo {volume} {11}},\
  \bibinfo {pages} {330} (\bibinfo {year} {2016})}\BibitemShut {NoStop}%
\bibitem [{\citenamefont {Elzerman}\ \emph {et~al.}(2004)\citenamefont
  {Elzerman}, \citenamefont {Hanson}, \citenamefont {Willems~van Beveren},
  \citenamefont {Witkamp}, \citenamefont {Vandersypen},\ and\ \citenamefont
  {Kouwenhoven}}]{Elzerman2004}%
  \BibitemOpen
  \bibfield  {author} {\bibinfo {author} {\bibfnamefont {J.~M.}\ \bibnamefont
  {Elzerman}}, \bibinfo {author} {\bibfnamefont {R.}~\bibnamefont {Hanson}},
  \bibinfo {author} {\bibfnamefont {L.~H.}\ \bibnamefont {Willems~van
  Beveren}}, \bibinfo {author} {\bibfnamefont {B.}~\bibnamefont {Witkamp}},
  \bibinfo {author} {\bibfnamefont {L.~M.~K.}\ \bibnamefont {Vandersypen}}, \
  and\ \bibinfo {author} {\bibfnamefont {L.~P.}\ \bibnamefont {Kouwenhoven}},\
  }\href {http://dx.doi.org/10.1038/nature02693} {\bibfield  {journal}
  {\bibinfo  {journal} {Nature}\ }\textbf {\bibinfo {volume} {430}},\ \bibinfo
  {pages} {431} (\bibinfo {year} {2004})}\BibitemShut {NoStop}%
\bibitem [{\citenamefont {Petta}\ \emph {et~al.}(2005)\citenamefont {Petta},
  \citenamefont {Johnson}, \citenamefont {Taylor}, \citenamefont {Laird},
  \citenamefont {Yacoby}, \citenamefont {Lukin}, \citenamefont {Marcus},
  \citenamefont {Hanson},\ and\ \citenamefont {Gossard}}]{Petta2180}%
  \BibitemOpen
  \bibfield  {author} {\bibinfo {author} {\bibfnamefont {J.~R.}\ \bibnamefont
  {Petta}}, \bibinfo {author} {\bibfnamefont {A.~C.}\ \bibnamefont {Johnson}},
  \bibinfo {author} {\bibfnamefont {J.~M.}\ \bibnamefont {Taylor}}, \bibinfo
  {author} {\bibfnamefont {E.~A.}\ \bibnamefont {Laird}}, \bibinfo {author}
  {\bibfnamefont {A.}~\bibnamefont {Yacoby}}, \bibinfo {author} {\bibfnamefont
  {M.~D.}\ \bibnamefont {Lukin}}, \bibinfo {author} {\bibfnamefont {C.~M.}\
  \bibnamefont {Marcus}}, \bibinfo {author} {\bibfnamefont {M.~P.}\
  \bibnamefont {Hanson}}, \ and\ \bibinfo {author} {\bibfnamefont {A.~C.}\
  \bibnamefont {Gossard}},\ }\href {\doibase 10.1126/science.1116955}
  {\bibfield  {journal} {\bibinfo  {journal} {Science}\ }\textbf {\bibinfo
  {volume} {309}},\ \bibinfo {pages} {2180} (\bibinfo {year}
  {2005})}\BibitemShut {NoStop}%
\bibitem [{\citenamefont {Pla}\ \emph {et~al.}(2012)\citenamefont {Pla},
  \citenamefont {Tan}, \citenamefont {Dehollain}, \citenamefont {Lim},
  \citenamefont {Morton}, \citenamefont {Jamieson}, \citenamefont {Dzurak},\
  and\ \citenamefont {Morello}}]{Pla2012}%
  \BibitemOpen
  \bibfield  {author} {\bibinfo {author} {\bibfnamefont {J.~J.}\ \bibnamefont
  {Pla}}, \bibinfo {author} {\bibfnamefont {K.~Y.}\ \bibnamefont {Tan}},
  \bibinfo {author} {\bibfnamefont {J.~P.}\ \bibnamefont {Dehollain}}, \bibinfo
  {author} {\bibfnamefont {W.~H.}\ \bibnamefont {Lim}}, \bibinfo {author}
  {\bibfnamefont {J.~J.}\ \bibnamefont {Morton}}, \bibinfo {author}
  {\bibfnamefont {D.~N.}\ \bibnamefont {Jamieson}}, \bibinfo {author}
  {\bibfnamefont {A.~S.}\ \bibnamefont {Dzurak}}, \ and\ \bibinfo {author}
  {\bibfnamefont {A.}~\bibnamefont {Morello}},\ }\href
  {https://www.nature.com/articles/nature11449} {\bibfield  {journal} {\bibinfo
   {journal} {Nature}\ }\textbf {\bibinfo {volume} {489}},\ \bibinfo {pages}
  {541} (\bibinfo {year} {2012})}\BibitemShut {NoStop}%
\bibitem [{\citenamefont {Pioro-Ladriere}\ \emph {et~al.}(2008)\citenamefont
  {Pioro-Ladriere}, \citenamefont {Obata}, \citenamefont {Tokura},
  \citenamefont {Shin}, \citenamefont {Kubo}, \citenamefont {Yoshida},
  \citenamefont {Taniyama},\ and\ \citenamefont {Tarucha}}]{Pioro2008}%
  \BibitemOpen
  \bibfield  {author} {\bibinfo {author} {\bibfnamefont {M.}~\bibnamefont
  {Pioro-Ladriere}}, \bibinfo {author} {\bibfnamefont {T.}~\bibnamefont
  {Obata}}, \bibinfo {author} {\bibfnamefont {Y.}~\bibnamefont {Tokura}},
  \bibinfo {author} {\bibfnamefont {Y.-S.}\ \bibnamefont {Shin}}, \bibinfo
  {author} {\bibfnamefont {T.}~\bibnamefont {Kubo}}, \bibinfo {author}
  {\bibfnamefont {K.}~\bibnamefont {Yoshida}}, \bibinfo {author} {\bibfnamefont
  {T.}~\bibnamefont {Taniyama}}, \ and\ \bibinfo {author} {\bibfnamefont
  {S.}~\bibnamefont {Tarucha}},\ }\href
  {https://www.nature.com/articles/nphys1053} {\bibfield  {journal} {\bibinfo
  {journal} {Nature Physics}\ }\textbf {\bibinfo {volume} {4}},\ \bibinfo
  {pages} {776} (\bibinfo {year} {2008})}\BibitemShut {NoStop}%
\bibitem [{\citenamefont {Maurand}\ \emph {et~al.}(2016)\citenamefont
  {Maurand}, \citenamefont {Jehl}, \citenamefont {Kotekar-Patil}, \citenamefont
  {Corna}, \citenamefont {Bohuslavskyi}, \citenamefont {Lavi{\'e}ville},
  \citenamefont {Hutin}, \citenamefont {Barraud}, \citenamefont {Vinet},
  \citenamefont {Sanquer} \emph {et~al.}}]{Maurand2016}%
  \BibitemOpen
  \bibfield  {author} {\bibinfo {author} {\bibfnamefont {R.}~\bibnamefont
  {Maurand}}, \bibinfo {author} {\bibfnamefont {X.}~\bibnamefont {Jehl}},
  \bibinfo {author} {\bibfnamefont {D.}~\bibnamefont {Kotekar-Patil}}, \bibinfo
  {author} {\bibfnamefont {A.}~\bibnamefont {Corna}}, \bibinfo {author}
  {\bibfnamefont {H.}~\bibnamefont {Bohuslavskyi}}, \bibinfo {author}
  {\bibfnamefont {R.}~\bibnamefont {Lavi{\'e}ville}}, \bibinfo {author}
  {\bibfnamefont {L.}~\bibnamefont {Hutin}}, \bibinfo {author} {\bibfnamefont
  {S.}~\bibnamefont {Barraud}}, \bibinfo {author} {\bibfnamefont
  {M.}~\bibnamefont {Vinet}}, \bibinfo {author} {\bibfnamefont
  {M.}~\bibnamefont {Sanquer}},  \emph {et~al.},\ }\href
  {https://www.nature.com/articles/ncomms13575} {\bibfield  {journal} {\bibinfo
   {journal} {Nature Communications}\ }\textbf {\bibinfo {volume} {7}},\
  \bibinfo {pages} {13575} (\bibinfo {year} {2016})}\BibitemShut {NoStop}%
\bibitem [{\citenamefont {Huang}\ \emph {et~al.}(2017)\citenamefont {Huang},
  \citenamefont {Veldhorst}, \citenamefont {Zimmerman}, \citenamefont
  {Dzurak},\ and\ \citenamefont {Culcer}}]{Huang2017}%
  \BibitemOpen
  \bibfield  {author} {\bibinfo {author} {\bibfnamefont {W.}~\bibnamefont
  {Huang}}, \bibinfo {author} {\bibfnamefont {M.}~\bibnamefont {Veldhorst}},
  \bibinfo {author} {\bibfnamefont {N.~M.}\ \bibnamefont {Zimmerman}}, \bibinfo
  {author} {\bibfnamefont {A.~S.}\ \bibnamefont {Dzurak}}, \ and\ \bibinfo
  {author} {\bibfnamefont {D.}~\bibnamefont {Culcer}},\ }\href {\doibase
  10.1103/PhysRevB.95.075403} {\bibfield  {journal} {\bibinfo  {journal} {Phys.
  Rev. B}\ }\textbf {\bibinfo {volume} {95}},\ \bibinfo {pages} {075403}
  (\bibinfo {year} {2017})}\BibitemShut {NoStop}%
\bibitem [{\citenamefont {Corna}\ \emph {et~al.}(2018)\citenamefont {Corna},
  \citenamefont {Bourdet}, \citenamefont {Maurand}, \citenamefont {Crippa},
  \citenamefont {Kotekar-Patil}, \citenamefont {Bohuslavskyi}, \citenamefont
  {Lavi{\'e}ville}, \citenamefont {Hutin}, \citenamefont {Barraud},
  \citenamefont {Jehl}, \citenamefont {Vinet}, \citenamefont {De~Franceschi},
  \citenamefont {Niquet},\ and\ \citenamefont {Sanquer}}]{Corna2018}%
  \BibitemOpen
  \bibfield  {author} {\bibinfo {author} {\bibfnamefont {A.}~\bibnamefont
  {Corna}}, \bibinfo {author} {\bibfnamefont {L.}~\bibnamefont {Bourdet}},
  \bibinfo {author} {\bibfnamefont {R.}~\bibnamefont {Maurand}}, \bibinfo
  {author} {\bibfnamefont {A.}~\bibnamefont {Crippa}}, \bibinfo {author}
  {\bibfnamefont {D.}~\bibnamefont {Kotekar-Patil}}, \bibinfo {author}
  {\bibfnamefont {H.}~\bibnamefont {Bohuslavskyi}}, \bibinfo {author}
  {\bibfnamefont {R.}~\bibnamefont {Lavi{\'e}ville}}, \bibinfo {author}
  {\bibfnamefont {L.}~\bibnamefont {Hutin}}, \bibinfo {author} {\bibfnamefont
  {S.}~\bibnamefont {Barraud}}, \bibinfo {author} {\bibfnamefont
  {X.}~\bibnamefont {Jehl}}, \bibinfo {author} {\bibfnamefont {M.}~\bibnamefont
  {Vinet}}, \bibinfo {author} {\bibfnamefont {S.}~\bibnamefont
  {De~Franceschi}}, \bibinfo {author} {\bibfnamefont {Y.-M.}\ \bibnamefont
  {Niquet}}, \ and\ \bibinfo {author} {\bibfnamefont {M.}~\bibnamefont
  {Sanquer}},\ }\href {\doibase 10.1038/s41534-018-0059-1} {\bibfield
  {journal} {\bibinfo  {journal} {npj Quantum Information}\ }\textbf {\bibinfo
  {volume} {4}},\ \bibinfo {pages} {6} (\bibinfo {year} {2018})}\BibitemShut
  {NoStop}%
\bibitem [{\citenamefont {Muhonen}\ \emph {et~al.}(2015)\citenamefont
  {Muhonen}, \citenamefont {Laucht}, \citenamefont {Simmons}, \citenamefont
  {Dehollain}, \citenamefont {Kalra}, \citenamefont {Hudson}, \citenamefont
  {Freer}, \citenamefont {Itoh}, \citenamefont {Jamieson}, \citenamefont
  {McCallum} \emph {et~al.}}]{Muhonen2015}%
  \BibitemOpen
  \bibfield  {author} {\bibinfo {author} {\bibfnamefont {J.~T.}\ \bibnamefont
  {Muhonen}}, \bibinfo {author} {\bibfnamefont {A.}~\bibnamefont {Laucht}},
  \bibinfo {author} {\bibfnamefont {S.}~\bibnamefont {Simmons}}, \bibinfo
  {author} {\bibfnamefont {J.~P.}\ \bibnamefont {Dehollain}}, \bibinfo {author}
  {\bibfnamefont {R.}~\bibnamefont {Kalra}}, \bibinfo {author} {\bibfnamefont
  {F.~E.}\ \bibnamefont {Hudson}}, \bibinfo {author} {\bibfnamefont
  {S.}~\bibnamefont {Freer}}, \bibinfo {author} {\bibfnamefont {K.~M.}\
  \bibnamefont {Itoh}}, \bibinfo {author} {\bibfnamefont {D.~N.}\ \bibnamefont
  {Jamieson}}, \bibinfo {author} {\bibfnamefont {J.~C.}\ \bibnamefont
  {McCallum}},  \emph {et~al.},\ }\href
  {http://iopscience.iop.org/article/10.1088/0953-8984/27/15/154205/meta}
  {\bibfield  {journal} {\bibinfo  {journal} {Journal of Physics: Condensed
  Matter}\ }\textbf {\bibinfo {volume} {27}},\ \bibinfo {pages} {154205}
  (\bibinfo {year} {2015})}\BibitemShut {NoStop}%
\bibitem [{\citenamefont {Chan}\ \emph {et~al.}(2018)\citenamefont {Chan},
  \citenamefont {Huang}, \citenamefont {Yang}, \citenamefont {Hwang},
  \citenamefont {Hensen}, \citenamefont {Tanttu}, \citenamefont {Hudson},
  \citenamefont {Itoh}, \citenamefont {Laucht}, \citenamefont {Morello} \emph
  {et~al.}}]{Chan2018}%
  \BibitemOpen
  \bibfield  {author} {\bibinfo {author} {\bibfnamefont {K.}~\bibnamefont
  {Chan}}, \bibinfo {author} {\bibfnamefont {W.}~\bibnamefont {Huang}},
  \bibinfo {author} {\bibfnamefont {C.}~\bibnamefont {Yang}}, \bibinfo {author}
  {\bibfnamefont {J.}~\bibnamefont {Hwang}}, \bibinfo {author} {\bibfnamefont
  {B.}~\bibnamefont {Hensen}}, \bibinfo {author} {\bibfnamefont
  {T.}~\bibnamefont {Tanttu}}, \bibinfo {author} {\bibfnamefont
  {F.}~\bibnamefont {Hudson}}, \bibinfo {author} {\bibfnamefont
  {K.}~\bibnamefont {Itoh}}, \bibinfo {author} {\bibfnamefont {A.}~\bibnamefont
  {Laucht}}, \bibinfo {author} {\bibfnamefont {A.}~\bibnamefont {Morello}},
  \emph {et~al.},\ }\href {https://arxiv.org/abs/1803.01609} {\bibfield
  {journal} {\bibinfo  {journal} {arXiv:1803.01609}\ } (\bibinfo {year}
  {2018})}\BibitemShut {NoStop}%
\bibitem [{\citenamefont {Nowack}\ \emph {et~al.}(2011)\citenamefont {Nowack},
  \citenamefont {Shafiei}, \citenamefont {Laforest}, \citenamefont
  {Prawiroatmodjo}, \citenamefont {Schreiber}, \citenamefont {Reichl},
  \citenamefont {Wegscheider},\ and\ \citenamefont {Vandersypen}}]{Nowack1269}%
  \BibitemOpen
  \bibfield  {author} {\bibinfo {author} {\bibfnamefont {K.~C.}\ \bibnamefont
  {Nowack}}, \bibinfo {author} {\bibfnamefont {M.}~\bibnamefont {Shafiei}},
  \bibinfo {author} {\bibfnamefont {M.}~\bibnamefont {Laforest}}, \bibinfo
  {author} {\bibfnamefont {G.~E. D.~K.}\ \bibnamefont {Prawiroatmodjo}},
  \bibinfo {author} {\bibfnamefont {L.~R.}\ \bibnamefont {Schreiber}}, \bibinfo
  {author} {\bibfnamefont {C.}~\bibnamefont {Reichl}}, \bibinfo {author}
  {\bibfnamefont {W.}~\bibnamefont {Wegscheider}}, \ and\ \bibinfo {author}
  {\bibfnamefont {L.~M.~K.}\ \bibnamefont {Vandersypen}},\ }\href {\doibase
  10.1126/science.1209524} {\bibfield  {journal} {\bibinfo  {journal}
  {Science}\ }\textbf {\bibinfo {volume} {333}},\ \bibinfo {pages} {1269}
  (\bibinfo {year} {2011})}\BibitemShut {NoStop}%
\bibitem [{\citenamefont {Kalra}\ \emph {et~al.}(2014)\citenamefont {Kalra},
  \citenamefont {Laucht}, \citenamefont {Hill},\ and\ \citenamefont
  {Morello}}]{Kalra2014}%
  \BibitemOpen
  \bibfield  {author} {\bibinfo {author} {\bibfnamefont {R.}~\bibnamefont
  {Kalra}}, \bibinfo {author} {\bibfnamefont {A.}~\bibnamefont {Laucht}},
  \bibinfo {author} {\bibfnamefont {C.~D.}\ \bibnamefont {Hill}}, \ and\
  \bibinfo {author} {\bibfnamefont {A.}~\bibnamefont {Morello}},\ }\href
  {\doibase 10.1103/PhysRevX.4.021044} {\bibfield  {journal} {\bibinfo
  {journal} {Phys. Rev. X}\ }\textbf {\bibinfo {volume} {4}},\ \bibinfo {pages}
  {021044} (\bibinfo {year} {2014})}\BibitemShut {NoStop}%
\bibitem [{\citenamefont {Angus}\ \emph {et~al.}(2007)\citenamefont {Angus},
  \citenamefont {Ferguson}, \citenamefont {Dzurak},\ and\ \citenamefont
  {Clark}}]{Angus2007}%
  \BibitemOpen
  \bibfield  {author} {\bibinfo {author} {\bibfnamefont {S.~J.}\ \bibnamefont
  {Angus}}, \bibinfo {author} {\bibfnamefont {A.~J.}\ \bibnamefont {Ferguson}},
  \bibinfo {author} {\bibfnamefont {A.~S.}\ \bibnamefont {Dzurak}}, \ and\
  \bibinfo {author} {\bibfnamefont {R.~G.}\ \bibnamefont {Clark}},\ }\href
  {\doibase 10.1021/nl070949k} {\bibfield  {journal} {\bibinfo  {journal} {Nano
  Letters}\ }\textbf {\bibinfo {volume} {7}},\ \bibinfo {pages} {2051}
  (\bibinfo {year} {2007})}\BibitemShut {NoStop}%
\bibitem [{\citenamefont {Thorbeck}\ and\ \citenamefont
  {Zimmerman}(2015)}]{Thorbeck2015}%
  \BibitemOpen
  \bibfield  {author} {\bibinfo {author} {\bibfnamefont {T.}~\bibnamefont
  {Thorbeck}}\ and\ \bibinfo {author} {\bibfnamefont {N.~M.}\ \bibnamefont
  {Zimmerman}},\ }\href {\doibase 10.1063/1.4928320} {\bibfield  {journal}
  {\bibinfo  {journal} {AIP Advances}\ }\textbf {\bibinfo {volume} {5}},\
  \bibinfo {pages} {087107} (\bibinfo {year} {2015})}\BibitemShut {NoStop}%
\bibitem [{\citenamefont {Dehollain}\ \emph {et~al.}(2012)\citenamefont
  {Dehollain}, \citenamefont {Pla}, \citenamefont {Siew}, \citenamefont {Tan},
  \citenamefont {Dzurak},\ and\ \citenamefont {Morello}}]{Dehollain2012}%
  \BibitemOpen
  \bibfield  {author} {\bibinfo {author} {\bibfnamefont {J.}~\bibnamefont
  {Dehollain}}, \bibinfo {author} {\bibfnamefont {J.}~\bibnamefont {Pla}},
  \bibinfo {author} {\bibfnamefont {E.}~\bibnamefont {Siew}}, \bibinfo {author}
  {\bibfnamefont {K.}~\bibnamefont {Tan}}, \bibinfo {author} {\bibfnamefont
  {A.}~\bibnamefont {Dzurak}}, \ and\ \bibinfo {author} {\bibfnamefont
  {A.}~\bibnamefont {Morello}},\ }\href
  {http://iopscience.iop.org/article/10.1088/0957-4484/24/1/015202/meta}
  {\bibfield  {journal} {\bibinfo  {journal} {Nanotechnology}\ }\textbf
  {\bibinfo {volume} {24}},\ \bibinfo {pages} {015202} (\bibinfo {year}
  {2012})}\BibitemShut {NoStop}%
\bibitem [{\citenamefont {Yang}\ \emph {et~al.}(2011)\citenamefont {Yang},
  \citenamefont {Lim}, \citenamefont {Zwanenburg},\ and\ \citenamefont
  {Dzurak}}]{Yang2011}%
  \BibitemOpen
  \bibfield  {author} {\bibinfo {author} {\bibfnamefont {C.~H.}\ \bibnamefont
  {Yang}}, \bibinfo {author} {\bibfnamefont {W.~H.}\ \bibnamefont {Lim}},
  \bibinfo {author} {\bibfnamefont {F.~A.}\ \bibnamefont {Zwanenburg}}, \ and\
  \bibinfo {author} {\bibfnamefont {A.~S.}\ \bibnamefont {Dzurak}},\ }\href
  {\doibase 10.1063/1.3654496} {\bibfield  {journal} {\bibinfo  {journal} {AIP
  Advances}\ }\textbf {\bibinfo {volume} {1}},\ \bibinfo {pages} {042111}
  (\bibinfo {year} {2011})}\BibitemShut {NoStop}%
\bibitem [{\citenamefont {Meunier}\ \emph {et~al.}(2011)\citenamefont
  {Meunier}, \citenamefont {Calado},\ and\ \citenamefont
  {Vandersypen}}]{Meunier2011}%
  \BibitemOpen
  \bibfield  {author} {\bibinfo {author} {\bibfnamefont {T.}~\bibnamefont
  {Meunier}}, \bibinfo {author} {\bibfnamefont {V.~E.}\ \bibnamefont {Calado}},
  \ and\ \bibinfo {author} {\bibfnamefont {L.~M.~K.}\ \bibnamefont
  {Vandersypen}},\ }\href {\doibase 10.1103/PhysRevB.83.121403} {\bibfield
  {journal} {\bibinfo  {journal} {Phys. Rev. B}\ }\textbf {\bibinfo {volume}
  {83}},\ \bibinfo {pages} {121403} (\bibinfo {year} {2011})}\BibitemShut
  {NoStop}%
\bibitem [{\citenamefont {Veldhorst}\ \emph
  {et~al.}(2015{\natexlab{b}})\citenamefont {Veldhorst}, \citenamefont
  {Ruskov}, \citenamefont {Yang}, \citenamefont {Hwang}, \citenamefont
  {Hudson}, \citenamefont {Flatt\'e}, \citenamefont {Tahan}, \citenamefont
  {Itoh}, \citenamefont {Morello},\ and\ \citenamefont
  {Dzurak}}]{Veldhorst2015b}%
  \BibitemOpen
  \bibfield  {author} {\bibinfo {author} {\bibfnamefont {M.}~\bibnamefont
  {Veldhorst}}, \bibinfo {author} {\bibfnamefont {R.}~\bibnamefont {Ruskov}},
  \bibinfo {author} {\bibfnamefont {C.~H.}\ \bibnamefont {Yang}}, \bibinfo
  {author} {\bibfnamefont {J.~C.~C.}\ \bibnamefont {Hwang}}, \bibinfo {author}
  {\bibfnamefont {F.~E.}\ \bibnamefont {Hudson}}, \bibinfo {author}
  {\bibfnamefont {M.~E.}\ \bibnamefont {Flatt\'e}}, \bibinfo {author}
  {\bibfnamefont {C.}~\bibnamefont {Tahan}}, \bibinfo {author} {\bibfnamefont
  {K.~M.}\ \bibnamefont {Itoh}}, \bibinfo {author} {\bibfnamefont
  {A.}~\bibnamefont {Morello}}, \ and\ \bibinfo {author} {\bibfnamefont
  {A.~S.}\ \bibnamefont {Dzurak}},\ }\href {\doibase
  10.1103/PhysRevB.92.201401} {\bibfield  {journal} {\bibinfo  {journal} {Phys.
  Rev. B}\ }\textbf {\bibinfo {volume} {92}},\ \bibinfo {pages} {201401}
  (\bibinfo {year} {2015}{\natexlab{b}})}\BibitemShut {NoStop}%
\bibitem [{\citenamefont {McKay}\ \emph {et~al.}(2017)\citenamefont {McKay},
  \citenamefont {Wood}, \citenamefont {Sheldon}, \citenamefont {Chow},\ and\
  \citenamefont {Gambetta}}]{PhysRevA.96.022330}%
  \BibitemOpen
  \bibfield  {author} {\bibinfo {author} {\bibfnamefont {D.~C.}\ \bibnamefont
  {McKay}}, \bibinfo {author} {\bibfnamefont {C.~J.}\ \bibnamefont {Wood}},
  \bibinfo {author} {\bibfnamefont {S.}~\bibnamefont {Sheldon}}, \bibinfo
  {author} {\bibfnamefont {J.~M.}\ \bibnamefont {Chow}}, \ and\ \bibinfo
  {author} {\bibfnamefont {J.~M.}\ \bibnamefont {Gambetta}},\ }\href {\doibase
  10.1103/PhysRevA.96.022330} {\bibfield  {journal} {\bibinfo  {journal} {Phys.
  Rev. A}\ }\textbf {\bibinfo {volume} {96}},\ \bibinfo {pages} {022330}
  (\bibinfo {year} {2017})}\BibitemShut {NoStop}%
\bibitem [{\citenamefont {Ryan}\ \emph {et~al.}(2009)\citenamefont {Ryan},
  \citenamefont {Laforest},\ and\ \citenamefont {Laflamme}}]{Ryan2009}%
  \BibitemOpen
  \bibfield  {author} {\bibinfo {author} {\bibfnamefont {C.}~\bibnamefont
  {Ryan}}, \bibinfo {author} {\bibfnamefont {M.}~\bibnamefont {Laforest}}, \
  and\ \bibinfo {author} {\bibfnamefont {R.}~\bibnamefont {Laflamme}},\ }\href
  {http://iopscience.iop.org/article/10.1088/1367-2630/11/1/013034/meta}
  {\bibfield  {journal} {\bibinfo  {journal} {New Journal of Physics}\ }\textbf
  {\bibinfo {volume} {11}},\ \bibinfo {pages} {013034} (\bibinfo {year}
  {2009})}\BibitemShut {NoStop}%
\bibitem [{\citenamefont {Laucht}\ \emph {et~al.}(2017)\citenamefont {Laucht},
  \citenamefont {Kalra}, \citenamefont {Simmons}, \citenamefont {Dehollain},
  \citenamefont {Muhonen}, \citenamefont {Mohiyaddin}, \citenamefont {Freer},
  \citenamefont {Hudson}, \citenamefont {Itoh}, \citenamefont {Jamieson} \emph
  {et~al.}}]{Laucht2017}%
  \BibitemOpen
  \bibfield  {author} {\bibinfo {author} {\bibfnamefont {A.}~\bibnamefont
  {Laucht}}, \bibinfo {author} {\bibfnamefont {R.}~\bibnamefont {Kalra}},
  \bibinfo {author} {\bibfnamefont {S.}~\bibnamefont {Simmons}}, \bibinfo
  {author} {\bibfnamefont {J.~P.}\ \bibnamefont {Dehollain}}, \bibinfo {author}
  {\bibfnamefont {J.~T.}\ \bibnamefont {Muhonen}}, \bibinfo {author}
  {\bibfnamefont {F.~A.}\ \bibnamefont {Mohiyaddin}}, \bibinfo {author}
  {\bibfnamefont {S.}~\bibnamefont {Freer}}, \bibinfo {author} {\bibfnamefont
  {F.~E.}\ \bibnamefont {Hudson}}, \bibinfo {author} {\bibfnamefont {K.~M.}\
  \bibnamefont {Itoh}}, \bibinfo {author} {\bibfnamefont {D.~N.}\ \bibnamefont
  {Jamieson}},  \emph {et~al.},\ }\href
  {https://www.nature.com/articles/nnano.2016.178} {\bibfield  {journal}
  {\bibinfo  {journal} {Nature Nanotechnology}\ }\textbf {\bibinfo {volume}
  {12}},\ \bibinfo {pages} {61} (\bibinfo {year} {2017})}\BibitemShut {NoStop}%
\bibitem [{\citenamefont {Sergeevich}\ \emph {et~al.}(2011)\citenamefont
  {Sergeevich}, \citenamefont {Chandran}, \citenamefont {Combes}, \citenamefont
  {Bartlett},\ and\ \citenamefont {Wiseman}}]{Sergeevich2011}%
  \BibitemOpen
  \bibfield  {author} {\bibinfo {author} {\bibfnamefont {A.}~\bibnamefont
  {Sergeevich}}, \bibinfo {author} {\bibfnamefont {A.}~\bibnamefont
  {Chandran}}, \bibinfo {author} {\bibfnamefont {J.}~\bibnamefont {Combes}},
  \bibinfo {author} {\bibfnamefont {S.~D.}\ \bibnamefont {Bartlett}}, \ and\
  \bibinfo {author} {\bibfnamefont {H.~M.}\ \bibnamefont {Wiseman}},\ }\href
  {\doibase 10.1103/PhysRevA.84.052315} {\bibfield  {journal} {\bibinfo
  {journal} {Phys. Rev. A}\ }\textbf {\bibinfo {volume} {84}},\ \bibinfo
  {pages} {052315} (\bibinfo {year} {2011})}\BibitemShut {NoStop}%
\bibitem [{\citenamefont {Shulman}\ \emph {et~al.}(2014)\citenamefont
  {Shulman}, \citenamefont {Harvey}, \citenamefont {Nichol}, \citenamefont
  {Bartlett}, \citenamefont {Doherty}, \citenamefont {Umansky},\ and\
  \citenamefont {Yacoby}}]{Shulman2014}%
  \BibitemOpen
  \bibfield  {author} {\bibinfo {author} {\bibfnamefont {M.~D.}\ \bibnamefont
  {Shulman}}, \bibinfo {author} {\bibfnamefont {S.~P.}\ \bibnamefont {Harvey}},
  \bibinfo {author} {\bibfnamefont {J.~M.}\ \bibnamefont {Nichol}}, \bibinfo
  {author} {\bibfnamefont {S.~D.}\ \bibnamefont {Bartlett}}, \bibinfo {author}
  {\bibfnamefont {A.~C.}\ \bibnamefont {Doherty}}, \bibinfo {author}
  {\bibfnamefont {V.}~\bibnamefont {Umansky}}, \ and\ \bibinfo {author}
  {\bibfnamefont {A.}~\bibnamefont {Yacoby}},\ }\href
  {https://www.nature.com/articles/ncomms6156} {\bibfield  {journal} {\bibinfo
  {journal} {Nature Communications}\ }\textbf {\bibinfo {volume} {5}},\
  \bibinfo {pages} {5156} (\bibinfo {year} {2014})}\BibitemShut {NoStop}%
\bibitem [{\citenamefont {Delbecq}\ \emph {et~al.}(2016)\citenamefont
  {Delbecq}, \citenamefont {Nakajima}, \citenamefont {Stano}, \citenamefont
  {Otsuka}, \citenamefont {Amaha}, \citenamefont {Yoneda}, \citenamefont
  {Takeda}, \citenamefont {Allison}, \citenamefont {Ludwig}, \citenamefont
  {Wieck},\ and\ \citenamefont {Tarucha}}]{Delbecq2016}%
  \BibitemOpen
  \bibfield  {author} {\bibinfo {author} {\bibfnamefont {M.~R.}\ \bibnamefont
  {Delbecq}}, \bibinfo {author} {\bibfnamefont {T.}~\bibnamefont {Nakajima}},
  \bibinfo {author} {\bibfnamefont {P.}~\bibnamefont {Stano}}, \bibinfo
  {author} {\bibfnamefont {T.}~\bibnamefont {Otsuka}}, \bibinfo {author}
  {\bibfnamefont {S.}~\bibnamefont {Amaha}}, \bibinfo {author} {\bibfnamefont
  {J.}~\bibnamefont {Yoneda}}, \bibinfo {author} {\bibfnamefont
  {K.}~\bibnamefont {Takeda}}, \bibinfo {author} {\bibfnamefont
  {G.}~\bibnamefont {Allison}}, \bibinfo {author} {\bibfnamefont
  {A.}~\bibnamefont {Ludwig}}, \bibinfo {author} {\bibfnamefont {A.~D.}\
  \bibnamefont {Wieck}}, \ and\ \bibinfo {author} {\bibfnamefont
  {S.}~\bibnamefont {Tarucha}},\ }\href {\doibase
  10.1103/PhysRevLett.116.046802} {\bibfield  {journal} {\bibinfo  {journal}
  {Phys. Rev. Lett.}\ }\textbf {\bibinfo {volume} {116}},\ \bibinfo {pages}
  {046802} (\bibinfo {year} {2016})}\BibitemShut {NoStop}%
\bibitem [{\citenamefont {Vandersypen}\ \emph {et~al.}(2017)\citenamefont
  {Vandersypen}, \citenamefont {Bluhm}, \citenamefont {Clarke}, \citenamefont
  {Dzurak}, \citenamefont {Ishihara}, \citenamefont {Morello}, \citenamefont
  {Reilly}, \citenamefont {Schreiber},\ and\ \citenamefont
  {Veldhorst}}]{Vandersypen2017}%
  \BibitemOpen
  \bibfield  {author} {\bibinfo {author} {\bibfnamefont {L.}~\bibnamefont
  {Vandersypen}}, \bibinfo {author} {\bibfnamefont {H.}~\bibnamefont {Bluhm}},
  \bibinfo {author} {\bibfnamefont {J.}~\bibnamefont {Clarke}}, \bibinfo
  {author} {\bibfnamefont {A.}~\bibnamefont {Dzurak}}, \bibinfo {author}
  {\bibfnamefont {R.}~\bibnamefont {Ishihara}}, \bibinfo {author}
  {\bibfnamefont {A.}~\bibnamefont {Morello}}, \bibinfo {author} {\bibfnamefont
  {D.}~\bibnamefont {Reilly}}, \bibinfo {author} {\bibfnamefont
  {L.}~\bibnamefont {Schreiber}}, \ and\ \bibinfo {author} {\bibfnamefont
  {M.}~\bibnamefont {Veldhorst}},\ }\href
  {https://www.nature.com/articles/s41534-017-0038-y} {\bibfield  {journal}
  {\bibinfo  {journal} {npj Quantum Information}\ }\textbf {\bibinfo {volume}
  {3}},\ \bibinfo {pages} {34} (\bibinfo {year} {2017})}\BibitemShut {NoStop}%
\bibitem [{\citenamefont {Veldhorst}\ \emph {et~al.}(2017)\citenamefont
  {Veldhorst}, \citenamefont {Eenink}, \citenamefont {Yang},\ and\
  \citenamefont {Dzurak}}]{Veldhorst2017}%
  \BibitemOpen
  \bibfield  {author} {\bibinfo {author} {\bibfnamefont {M.}~\bibnamefont
  {Veldhorst}}, \bibinfo {author} {\bibfnamefont {H.}~\bibnamefont {Eenink}},
  \bibinfo {author} {\bibfnamefont {C.}~\bibnamefont {Yang}}, \ and\ \bibinfo
  {author} {\bibfnamefont {A.}~\bibnamefont {Dzurak}},\ }\href
  {https://www.nature.com/articles/s41467-017-01905-6} {\bibfield  {journal}
  {\bibinfo  {journal} {Nature Communications}\ }\textbf {\bibinfo {volume}
  {8}},\ \bibinfo {pages} {1766} (\bibinfo {year} {2017})}\BibitemShut
  {NoStop}%
\bibitem [{\citenamefont {Blume-Kohout}\ \emph {et~al.}(2017)\citenamefont
  {Blume-Kohout}, \citenamefont {Gamble}, \citenamefont {Nielsen},
  \citenamefont {Rudinger}, \citenamefont {Mizrahi}, \citenamefont {Fortier},\
  and\ \citenamefont {Maunz}}]{Blume-Kohout2017}%
  \BibitemOpen
  \bibfield  {author} {\bibinfo {author} {\bibfnamefont {R.}~\bibnamefont
  {Blume-Kohout}}, \bibinfo {author} {\bibfnamefont {J.~K.}\ \bibnamefont
  {Gamble}}, \bibinfo {author} {\bibfnamefont {E.}~\bibnamefont {Nielsen}},
  \bibinfo {author} {\bibfnamefont {K.}~\bibnamefont {Rudinger}}, \bibinfo
  {author} {\bibfnamefont {J.}~\bibnamefont {Mizrahi}}, \bibinfo {author}
  {\bibfnamefont {K.}~\bibnamefont {Fortier}}, \ and\ \bibinfo {author}
  {\bibfnamefont {P.}~\bibnamefont {Maunz}},\ }\href
  {http://dx.doi.org/10.1038/ncomms14485} {\bibfield  {journal} {\bibinfo
  {journal} {Nature Communications}\ }\textbf {\bibinfo {volume} {8}},\
  \bibinfo {pages} {14485} (\bibinfo {year} {2017})}\BibitemShut {NoStop}%
\end{thebibliography}%

\end{document}